\documentclass[aps,epsfig,twocolumn]{revtex4}

\usepackage{graphicx}
\usepackage{dcolumn}
\usepackage{bm}

\newcommand{\bq}{\begin{equation}}
\newcommand{\eq}{\end{equation}}
\newcommand{\bqa}{\begin{eqnarray}}
\newcommand{\eqa}{\end{eqnarray}}
\newcommand{\nn}{\nonumber \\}

\begin{document}
\draft
\title{ 
Emergent U(1) gauge theory with fractionalized boson/fermion 
from the bose condensation of exciton in multi-band insulator 
}

\author{Sung-Sik Lee and Patrick A. Lee}
\address{Department of Physics, Massachusetts Institute of Technology,\\
Cambridge, Massachusetts 02139, U.S.A.\\}
\date{\today}
       
\begin{abstract}
Fractionalized phases are studied in a low energy theory of exciton bose condensate in a multi-band insulator.
It is shown that U(1) gauge theory with either fractionalized boson or fermion can emerge out of a single model depending on the coupling constants.
Both the statistics and spin of the fractionalized particles are dynamically determined, 
satisfying the spin-statistics theorem in the continuum limit.
We present two mutually consistent descriptions for the fractionalization.
In the first approach, it is shown that fractionalized degree of freedom emerges
from reduced phase space constrained by strong interaction and
that the U(1) gauge field arises as a collective excitation of the low energy modes.
In the second approach, complimentary descriptions are provided for the fractionalization 
based on world line picture of the original excitons.
The emergent gauge structure is identified from the fluctuating web of exciton world lines 
which, in turn, realizes the string net condensation in a space-time picture.

\end{abstract}
\maketitle

\section{Introduction}
Fractionalization is a phenomenon 
where a microscopic degree of freedom 
in many-body system 
decays into multiple modes
by strong interaction.
For example, 
it is well known that interacting electrons in 1+1D
decay into spin and charge density waves\cite{HALDANE}. 
In 2+1D, electrons under strong magnetic field decay into 
quasiparticles carrying only fraction of the electron charge\cite{LAUGHLIN}.
Given the novelty of the phenomena, 
it is natural to ask whether the fractionalization
can occur more ubiquitously in other systems.

Possible fractionalized phases have been extensively studied
in 2+1D quantum spin system in view of its relevance to high temperature
superconductivity (a comprehensive review for this approach can be found in Ref.\cite{PALEE}).
Earlier, Anderson proposed that a neutral spin-$1/2$ excitation, dubbed as spinon,
can arise in a quantum disordered (spin liquid) phase of 
2+1D antiferromagnet\cite{ANDERSON}.
In gauge theory picture various mean-field states have been found 
which have the spinon excitation
along with a $Z_2$ gauge field\cite{READ1991,WEN1991,SENTHIL2000,WEN2002PRB}.
Since the $Z_2$ gauge field is discrete, 
the deconfinement phase can be stable with 
massive or massless spinon excitations.
If gauge group is continuous like U(1), 
the deconfinement phase with massive spinon
is not stable at zero temperature in 2+1D
owing to the proliferation of instantons\cite{POLYAKOV77}.
At zero temperature, the deconfinement phase can occur
if there is a time-reversal symmetry breaking\cite{WEN1989},
or there are massless modes\cite{IOFFE,SENTHIL04,HERMELE}.
The massless modes can occur at a
quantum critical point\cite{SENTHIL04} 
or they can occur more generically protected by
lattice symmetry\cite{WEN2002PRB,HERMELE}.
The existences of the deconfinement phase have been also demonstrated
in an exact soluble spin model\cite{KITAEV,WEN2003PRL} and 
a dimer model in frustrated lattice\cite{MOESSNER}.
More recently, various 3+1D models have been constructed
which show deconfinement phases\cite{WEN2002PRL,WEN2003}.
These models have played an important role 
in studying the existence of the fractionalized phase.
However they usually involve spin degrees of freedom 
defined on bond centers of lattices or four spin interaction which
break the rotation symmetry in spin space.
At the same time bosonic models have been constructed
which may be realized in principle in 
Josephson junction arrays\cite{MOTRUNICH,MOTRUNICH2004}.
It is clearly of interest to search for more examples 
of realistic models which support fractionalization.

In a sense the fractionalization is opposite to the phenomena where
multiple particles are bound to form a composite particle at low energy,
such as molecules made of atoms
and proton (neutron) made of quarks. 
The latter phenomena look more natural 
in the sense that larger structure is constructed  
from smaller building blocks.
Fractionalization 
seems to work in the opposite direction.
The puzzling nature of it becomes sharper when we realize 
that the fractionalization is about low energy physics.
Fractionalization concerns the emergence of atom-like objects out of `molecules'
at low energy, not probing real `atoms' inside `molecules'
at high energy.
This conceptual difficulty of identifying fractionalized degree of freedoms 
out of original particle translates into the difficulty in the formalism
to find a unique way of decomposition.
There is no unique way of assigning statistics and spin to
fractionalized particles.
For example, spin liquid phase can be described either by
bosonic spinon\cite{READ1991} or fermionic spinon\cite{WEN1991}.
Why should we choose one way of decomposition over another ?
More importantly, what is the fractionalized particle (emergent atom) after all,
if it is not the real constituent (real atom) inside the original particle (molecules) ? 

The purpose of the present paper is to answer these questions by 
demonstrating that not only the emergence of fractionalized particle 
but also the statistics and spin of the fractionalized
particle are all consequences of the intrinsic dynamics of the model.
We explicitly show that both fractionalized boson and fermion can emerge out of
a single theory depending on coupling constants, and 
that the spin and charge of the fractionalized particles are also
uniquely determined from dynamical constraint. 
The microscopic system we consider is 
a model describing excitons
in multi-band insulator.

Since fractionalization is a rather unfamiliar concept 
in condensed matter physics, 
the question is often asked, 
how is it possible to visualize
the new particles and the gauge field
in terms of the original particle ?
In this paper we provide an answer by
constructing a world line representation
of the fractionalized particles and the
gauge field in terms of the world lines of 
the original particles, which are in our case excitons.
The world line (or duality) picture is complimentary
to the more common field theoretical derivation,
in that the confined state which is
difficult to treat in field theory language,
is easier to visualize.
Furthermore, a qualitative understanding of the
transition between confined and deconfined phases
is possible in terms of pictures.
It is our hope that by explicitly working out
this example using both methods, we can
gain deeper insights into these fascinating
new phenomena.

We consider a system of spinless fermions with $N$ degenerate bands where 
each band is composed of a conduction band and a valence band.
The Hamiltonian is
\bqa
H & = &
\sum_{a=1}^{N}
\sum_k
\left[
  ( \epsilon^c_k - \mu )c^{a\dagger}_k c^a_k
+ ( \epsilon^d_k - \mu )d^{a\dagger}_k d^a_k
\right] \nn
&& + U \sum_{a=1}^{N} \sum_k ( c^{a\dagger}_k d^a_k + d^{a\dagger}_k  c^a_k ) \nn
&& - \frac{1}{ v} \sum_{a,b} \sum_{k,k^{'},q} 
V^{ab}_q 
c^{a\dagger}_{k+q} c^a_k d^b_{k^{'}+q} d^{b\dagger}_{k^{'}}.
\label{micro}
\eqa
Here $c^a_k$ ($d^a_k$) is the annihilation operator of the fermion 
in the a-th conduction (valence) band,
$\epsilon^c_k$ ($\epsilon^d_k$), the energy dispersion of the conduction (valence) band.
$\mu$ is the chemical potential.
For simplicity we set 
$\epsilon^c_k - \mu = \Delta^c + \frac{k^2}{2m^c}$ 
and $\epsilon^d_k - \mu = -\Delta^d - \frac{k^2}{2m^d}$.
The band gap is $\Delta^c + \Delta^c$ and $\mu$ is tuned so that
there is equal density of conduction  particle
and valence hole.
The index $a$ labels the band degeneracy which is $N$
for both conduction and valence bands.
We imagine that the index labels orbital degeneracy,
but we shall refer to it as the flavor index
following the particle physics literature.
$U$ is the overlap integral which admixes the conduction and valence bands.
$V^{ab}_q$ is the strength of the Coulomb repulsion between fermions 
in band $a$ and band $b$.
It is assumed that the Coulomb interaction is local in real space and 
that the strength of the interaction depends on 
whether fermions are in a same band or in different bands, that is,
$V^{ab}_q =  V + V^{'} \delta_{ab} $.
$v = L^{D-1}$ is the volume of the system in $d$ space dimension $d = (D-1)$.
We single out the attractive interaction between conduction particle
and valence hole.
The rest of the Coulomb interaction is assumed to be
already included in the band structure determination
of $\Delta^c + \Delta^d$ by the Hartree-Fock approximation.

The attractive interaction
leads to a bound state between particle and hole which is an exciton.
If the interaction energy is smaller than band gap,
the exciton state lies between the gap.
Since it is metastable, the fermion and hole will eventually recombine in a finite time.
If the binding energy is large enough to overcome the band gap,
particles and holes are spontaneously created 
to form a bose condensate in the ground state.
We will consider the latter case.

\begin{figure}
        \includegraphics[height=7cm,width=7cm]{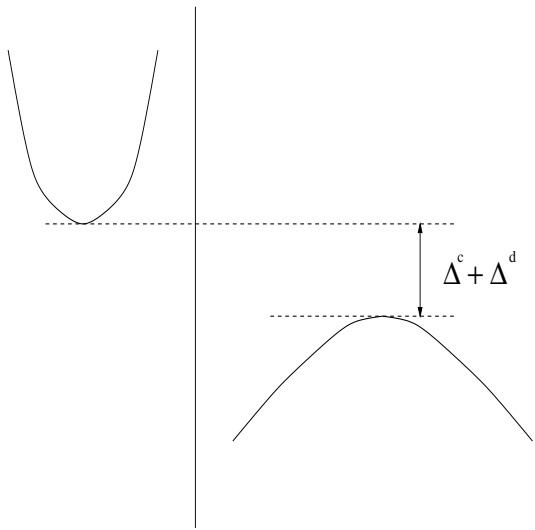}
\caption{
Schematic diagram of band structure in
the hetero-junction of semiconductors.
}
\label{fig:hetero}
\end{figure}

To fix ideas, it is useful to consider
one particular realization of this model.
Consider a hetero-junction where two
thin semiconductor layers are separated
by a thin barrier, shown schematically
in Fig. \ref{fig:hetero}.
The energy $\Delta^c + \Delta^d$ is the separation 
between the bottom of the conduction on the
left hand side and the top of the valence band 
on the right, and can in principle be tuned by
applying a voltage.
$U$ is the tunneling matrix element of 
an electron across the barrier.
In our model only electrons in bands with
the same orbital index are allowed to tunnel,
which is reasonable if we assume that 
different orbitals are orthogonal to each other
due to their symmetry within the plane.
If the Coulomb attraction between an electron on the left
and a hole on the right is strong, a finite
density of excitons will form spontaneously in the ground
state, resulting in a Bose condensate of excitons.
In real experiments it has proven difficult
to realize the Bose condensation of excitons in the bulk,
because the excitons tend to have attractive interaction,
leading to the formation of exciton molecules or
electron-hole liquid ground states.
The most promising current example is indeed the
layer geometry shown in Fig. \ref{fig:hetero} 
in GaAs hetero-junctions.
Instead of using a bias voltage, the conduction and valence
bands are populated by optical pumping,
leading to a meta-stable state\cite{BUTOV}.
The dipolar interaction between the excitons assures that
they are repulsive, making the system a promising candidate
for exciton condensation.
Here we are mostly interested in Eq. (\ref{micro}) as a model
system which we assume can be realized in both two 
and three dimensions.
The presence of $N$ degenerate bands lead to additional
complication in that there can be $N^2$ species of excitons
and some or all of them may or may not condense.
The order parameter for exciton condensation is 
\bq
\bar \chi^{ab} = < d_i^{b\dagger} c_i^a >,
\label{order}
\eq
where $i$ is a site index.
In this paper we examine the case where
$\bar \chi^{aa} = \chi_d \neq 0$, 
i.e., the diagonal exciton are condensed.
While it is possible that the symmetry breaking
happens spontaneously, 
the presence of the $U$ term in Eq. (\ref{micro})
explicitly breaks the symmetry and assures
that $\bar \chi^{aa} \neq 0$.
( In the case where the condensation of diagonal
excitons occurs spontaneously,
we can set $U=0$ in our model. )
We also assume that there is a finite 
amplitude for off diagonal exciton condensation.
In appendix A we work out the mean field theory
to show that this assumption is locally
stable in some parameter range.
In the remainder of the paper we examine
the effect of phase fluctuations of
$\chi^{ab} = \chi_o e^{i \theta^{ab}}$,
to see whether the off-diagonal condensate
may be destroyed, and to see under what
circumstances a novel state of matter
with fractionalized particle and gauge field
may emerge.

\begin{figure}
        \includegraphics[height=10cm,width=9cm]{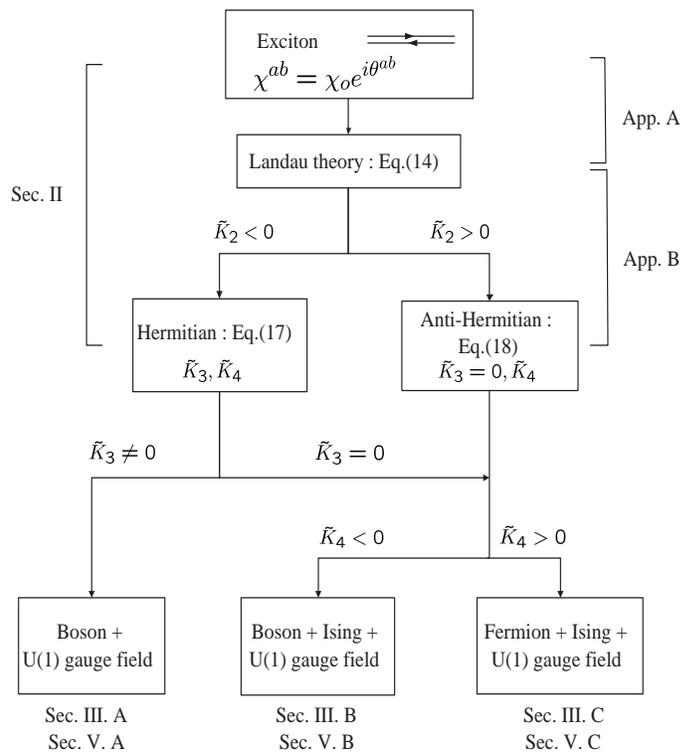}
\caption{
Flow chart from the microscopic exciton model
to low energy theories showing various fractionalized phases.
}
\label{fig:flow_chart}
\end{figure}

Fig. \ref{fig:flow_chart} provides a flow chart
for the development of this theory and can be used
as a road-map to this paper.
Starting from the microscopic Hamiltonian
given by Eq. (\ref{micro}) we can derive a 
Landau theory for the order parameter $\chi^{ab}$
based on symmetry considerations.
This is done in section II and the resulting
Landau theory is shown in Eq. (\ref{theaction}).
Since $\chi^{aa}$ are assumed to be condensed,
the remaining degrees of freedom are just
$\chi^{ab} = \chi_o e^{i \theta^{ab}}$ and
we focus only on phase fluctuations.
Note that $\chi^{ab}$ and $\chi^{ba}$ are 
in general independent because $\chi^{ab}$
is the pairing of conduction electron in
band $a$ with valence hole in band $b$, which is
distinguishable from pairing electron in
band $b$ and hole in band $a$.
Nevertheless, due to the condensation of $\chi^{aa}$,
there is a term which locks the phases,
which is proportional to
$\tilde K_2 \sum_{a \neq b}
\cos ( \theta^{ab}_i + \theta^{ba}_i)$.
For $\tilde K_2 < 0$, 
$ \theta^{ab}_i = - \theta^{ba}_i$ is favored.
Thus $\chi^{ab}$ is locked to $\chi^{ba*}$
and the Landau theory reduces to a Hermitian
matrix model Eq. (\ref{H}).
Similarly, for  $\tilde K_2 > 0$
$ \theta^{ab}_i = - \theta^{ba}_i + \pi$ 
is favored and we have the 
anti-Hermitian matrix model shown in 
Eq. (\ref{antiH}).
Further development depends on the next order
term in $\chi_o$.
For the Hermitian model, the third order term 
$ \tilde K_3 
\cos ( \theta^{ab}_i +\theta^{bc}_i +\theta^{ca}_i )$
is allowed which leads to the constraint
\bq
\theta^{ab}_i +\theta^{bc}_i +\theta^{ca}_i = 0
\label{intro_K3}
\eq
for strong coupling.
This constraint can be satisfied
by writing
$\theta^{ab}_i = \phi^a_i - \phi^b_i$.
The field $e^{i \phi^a_i}$
emerges as the boson field which carries
the flavor quantum number, but it is
coupled to a U(1) gauge field.
In $d=3$ and for large enough $N$,
a deconfined phase is possible.
It is called the Coulomb phase, because
it mimics our world of non-compact 
QED coupled to charged particles.
The Coulomb phase lies between the confined 
phase and the Higgs phase as a function of 
coupling constant $\kappa$ (see Fig. \ref{fig:phase} (a)).
The Higgs phase is the phase where the off diagonal
excitons are condensed and the confined phase is
one where the excitons are disordered by phase 
fluctuations, with exponentially decaying
correlation functions.
The Coulomb phase is the novel phase which exhibits gapped 
fractionalized bosons and gapless U(1) gauge field
which are analogs of photons.
This is discussed in section III.A.

In section V. A we discuss this model again
using the world-line representation.
It is natural that the off diagonal exciton
$\chi^{ab}$ is represented by a double line
with $a$, $b$ labels.
In the strong coupling limit, these excitons scatter
with each other and exchange partners rapidly
(see Figs. \ref{fig:vertices} and \ref{fig:vertices_double}).
Thus the individual exciton loses its identity
on a short scale $\xi_1$.
However, a single line carrying a flavor index may survive
over a long scale $\xi_2$, as shown in Fig. \ref{fig:web}.
These single lines represent the world lines of the 
fractionalized particles.
However, it is clear that these world lines
are not freely propagating, but connected to each
other via a web representing multiple scattering 
with other excitons.
The web of excitons turn out to show the
same dynamics as the world sheet swept out by
a line of electric flux in a compact U(1) gauge theory.
This is reviewed in section IV.
The size of the web gives rise to a third length
scale $\xi_3$ shown in Fig. \ref{fig:web}.
The divergence of $\xi_3$ signals the transition
between the confined phase and the Coulomb (deconfined)
phase.
Finally the divergence of $\xi_2$ leads to the
Higgs phase of the coupled boson-gauge field system,
which corresponds to the Bose condensation of 
the original off diagonal excitons.

Next we turn to the anti-Hermitian model,
where $\tilde K_3 = 0$ by symmetry.
The next most important term is 
$\tilde K_4 \cos(
\theta^{ab}_i 
+\theta^{bc}_i 
+\theta^{cd}_i 
+\theta^{da}_i  )$.
For $\tilde K_4 < 0$
the term favors 
\bq
\theta^{ab}_i +\theta^{bc}_i +\theta^{cd}_i +\theta^{da}_i = 0.
\label{intro_K4}
\eq
These constraints are less restrictive
than Eq. (\ref{intro_K3}).
However the set of equations (\ref{intro_K4})
are not independent and it can be shown that
they are satisfied by
\bq
\theta^{ab}_i = \phi^a_i - \phi^b_i + \frac{\pi}{2}(1+z_i),
\eq
where $z_i = \pm 1$ is an Ising degree of freedom
on each site.
The resulting theory consists of the boson
$e^{i \phi^a_i}$ coupled to a U(1) gauge field
plus an Ising degree of freedom which is decoupled
at low energies, as shown in section III.B.
The world line version is discussed in Sec. V. B.

Finally for $\tilde K_4 > 0$, the constraint
\bq
\theta^{ab}_i +\theta^{bc}_i +\theta^{cd}_i +\theta^{da}_i = \pi
\label{intro_K4_2}
\eq
can not be satisfied by any bosonic parameterization.
This leads to a frustration among the phases which
suppresses the condensation of off diagonal excitons.
However, a fermionic decomposition satisfy the
constraint perfectly.
A U(1) gauge field is also generated, but it turns
out that a background $\pi$-flux is the favored saddle
point.
This leads to the familiar Dirac fermions centered
at $(\pm \frac{\pi}{2}, \pm  \frac{\pi}{2})$ in 2+1D.
The low energy fermion degrees of freedom is
represented by a four component Dirac field,
which can be interpreted as carrying 
spin degree of freedom, in agreement
with the spin-statistics theorem.
This is detailed in section III. C.
It turns out that the fermion also emerges from the 
world line picture, because we can show
that each closed loop of single world line
is associated with a minus sign, 
as is shown in section V. C.
However, the spinor nature of the fermion 
requires a description of the short distance physics
which is not easy to capture in the world line picture.

The readers who are interested only in 
the low energy physics of the phenomenological model
may skip the first part of Sec. II and start 
from the Landau model, (\ref{theaction}).
The readers who are familiar with the dual representation 
of the compact U(1) gauge theory may skip the Sec. IV.
Sec. III and Sec. V deal with the same phenomena in two complimentary approaches
.
Sec. III provides a more quantitative but abstract analysis 
of the phenomenon of fractionalization.
Sec. V is less quantitative but more physical insights can be obtained on what is really occurring 
in the original exciton language.

\section{Effective theory of exciton bose condensate}

Starting from Eq. (\ref{micro}), we treat
the attraction between the conduction electron
and the valence hole by a Hubbard-Stratonovich
decoupling.
The effective action for the exciton condensate and the fermions 
are obtained to be (for details, see Appendix A),
\bqa
S & = &  S_{1} + S_{2},
\label{ls}
\eqa
where
\bqa
S_{1} & = &
\int d\tau
\sum_k
\sum_a
\Bigl[
c^{a *}_k  \left( \partial_\tau + \epsilon^c_k -\mu \right) c_k^a  \nn
&& + d^{a *}_k \left( \partial_\tau +  \epsilon^d_k - \mu \right) d_k^a
\Bigr]  \nn
&& + V \sum_q 
\sum_{a,b}
\left(
 (\chi_q^\dagger)^{ab} \chi_q^{ba}
- \frac{1}{\sqrt{v}} \sum_{k} c^{a*}_{k+q} \chi_q^{ab}  d_{k}^b \right. \nn
&& \left. - \frac{1}{\sqrt{v}} \sum_{k} d^{b*}_{k} (\chi_q^\dagger)^{ba} c_{k+q}^{a}
\right),
\label{ls1}
\eqa
and
\bqa
S_{2} & = &
\int d\tau \left[
- \frac{V V^{'}}{V+V^{'}}  \sum_q \sum_{a} |\chi^{aa}_q|^2 \right. \nn
&& \left. + \sqrt{v} \frac{V U}{V+V^{'}} \sum_a ( \chi^{aa}_0 + \chi^{aa*}_0 ) \right].
\label{ls2}
\eqa
Here $\chi_k$ is the $N \times N$ complex matrix 
for the exciton condensate field in momentum space.
Each component $\chi^{ab}_k$ describes the condensate of excitons
made of particle in the $a$-th band and hole in the $b$-th band.
(Note that order of indices are important, $\chi^{ab} \neq \chi^{ba}$.)
Self-consistent equations are solved for $\chi^{ab}$.
Various solutions are possible depending on the values of the $V$, $V^{'}$ and $U$.
In Appendix A, a few saddle point solutions are obtained for $N=3$.
In the following we concentrate on one of the solutions, 
\bqa
\chi^{ab} & = & \chi_o + (\chi_d - \chi_o) \delta_{ab},
\label{an}
\eqa
where both the diagonal and off-diagonal components have finite amplitude.

With the exciton condensation the single particle excitation gap 
becomes larger than the bare band gap because it takes extra energy 
in breaking the particle-hole binding.
At energy far below the single particle excitation gap 
the fluctuations of the exciton bose condensates become effective degrees of freedom.
Integrating out the fermions we obtain the effective action for the exciton condensate fields $\chi$.
Symmetry enables us to construct the low energy theory without actual calculation.
Since the fermions appear only in $S_1$ the symmetry of $S_1$ determines
the form of the action generated by their integration.
$S_1$ is invariant under global $U(N) \times U(N)$  transformation,
\bqa
c_k & \rightarrow & {\cal U} c_k, \nn
d_k & \rightarrow & {\cal V} d_k, \nn
\chi_k & \rightarrow & {\cal U} \chi_k {\cal V}^\dagger,
\label{symmetry}
\eqa
where
$c_k^T = (c_k^1, c_k^2, ..., c_k^N)$,
$d_k^T = (d_k^1, d_k^2, ..., d_k^N)$
are N-component vectors for the fermion fields and
${\cal U}$ and ${\cal V}$ are independent $U(N)$ matrices.
The symmetry put stringent constraint on possible term in the effective action
because it must be invariant under the last line in Eq. (\ref{symmetry}).
Up to the sixth order we write the resulting effective action, 
\bqa
\tilde S_1   & = &
\sum_{{\bf i}} \int d \tau 
\Bigl[ 
k_\tau tr \chi_{{\bf i}}^\dagger \partial_\tau \chi_{{\bf i}} 
- k \sum_{\hat r = \hat 1,...,\hat d} tr ( \chi_{{\bf i}+\hat r}^\dagger \chi_{{\bf i}} + h.c.) \nn
&& + k_2  tr \chi_{{\bf i}}^\dagger \chi_{{\bf i}} + k_4^{'}  ( tr \chi_{{\bf i}}^\dagger \chi_{{\bf i}} )^2
+ k_4 tr \chi_{{\bf i}}^\dagger \chi_{{\bf i}} \chi_{{\bf i}}^\dagger \chi_{{\bf i}} \nn
&& + k_6  tr \chi_{{\bf i}}^\dagger \chi_{{\bf i}} \chi_{{\bf i}}^\dagger \chi_{{\bf i}} \chi_{{\bf i}}^\dagger \chi_{{\bf i}}
+ k_6^{'}  (tr \chi_{{\bf i}}^\dagger \chi_{{\bf i}}) ( tr \chi_{{\bf i}}^\dagger \chi_{{\bf i}} \chi_{{\bf i}}^\dagger \chi_{{\bf i}} ) \nn
&& + k_6^{''}  (tr \chi_{{\bf i}}^\dagger \chi_{{\bf i}})^3 \Bigr].
\eqa
Here $\chi_{\bf i}$ is the exciton condensate field at space coordinate ${\bf i}$.
$\hat r$ is a unit vector in the d-dimensional square (cubic) lattice for $d=2$ ($d=3$).
In principle, all of the parameters $k_\tau$, $k$ and $k_n$ can be determined 
from the Hamiltonian (\ref{micro}).
Note that there is only linear derivative in the imaginary time direction 
because the exciton is non-relativistic boson.
It is convenient to discretize the imaginary time direction to rewrite the above action as
\bqa
\tilde S_1 & = & \sum_i \Big[
- \frac{K}{2} tr \chi_{i+\Delta \tau}^\dagger \chi_{i}
- \frac{K}{4} \sum_{\hat r} tr ( \chi_{i+\hat r}^\dagger \chi_i + c.c.) \nn
&& + K_2  tr \chi_i^\dagger \chi_i + K_4^{'}  ( tr \chi_i^\dagger \chi_i )^2
+ K_4 tr \chi_i^\dagger \chi_i \chi_i^\dagger \chi_i \nn
&& + K_6  tr \chi_i^\dagger \chi_i \chi_i^\dagger \chi_i \chi_i^\dagger \chi_i
+ K_6^{'}  (tr \chi_i^\dagger \chi_i) ( tr \chi_i^\dagger \chi_i \chi_i^\dagger \chi_i ) \nn
&& + K_6^{''}  (tr \chi_i^\dagger \chi_i)^3
\Bigr].
\label{ds}
\eqa
Here $i = (\tau, {\bf i})$ is index for lattice point in D-dimensional Euclidean space-time.
The dimensionless coupling constants in the lattice action are given by
$K = 2 k_\tau$,
$K_2 = \Delta \tau k_2 + k_\tau$
and
$K_n = \Delta \tau k_n$, where
the discrete time step is set to be
$\Delta \tau = \frac{K}{4k}$.

The term $S_2$ in Eq. (\ref{ls2}) has lower symmetry.
The first term is proportional to $V^{'}$ which distinguishes
between the interaction between particles with the same and different
flavor indices.
The $U(N) \times U(N)$ symmetry is broken down to 
$U(1)^{2N}$ where each U(1) is generated by diagonal
$U(N)$ matrices ${\cal U}$ and ${\cal V}$, 
i.e., ${\cal U}_{ab} = \delta_{ab} e^{i \phi^a}$
and ${\cal V}_{ab} = \delta_{ab} e^{i \varphi^a}$.
The second term is proportional to $U$ which explicitly
break the phase symmetry of the diagonal exciton condensate.
This further breaks the symmetry down to $U(1)^N$ where each
$U(1)$ is generated by the diagonal matrix ${\cal U} = {\cal V}$.
In discrete space time the total action is easily written as
\bqa
\tilde S & = & \tilde S_1 + \tilde S_2, 
\label{theaction} \\
\tilde S_2 & = & \sum_i \Bigl[
- K_{V^{'}}   \sum_{a} |\chi^{aa}_i|^2
+ K_{U} \sum_a ( \chi^{aa}_i + \chi^{aa*}_i ) \Bigr], \nn
\eqa
where
$K_{V^{'}} =  \frac{V V^{'}}{V+V^{'}}  \Delta \tau$ and $K_U = \frac{V U}{V+V^{'}} \Delta \tau$.

The $U(1)^{N}$ transformation rotates only the phase of each component $\chi^{ab}$.
( Strictly speaking, the action (\ref{theaction}) has only $U(1)^{N-1}$ symmetry because
 $\chi$ itself is invariant under one U(1) transformation with ${\cal U}={\cal V} = e^{i\phi}$.)
Thus all amplitude fluctuations of $\chi^{ab}$ are gapped and we ignore
the amplitude fluctuations.
With the nonzero amplitudes for all exciton fields at the saddle point (\ref{an}),
we have $N^2$ independent U(1) phase modes.
The phases of the $N$ diagonal modes $\chi^{aa}$ are fixed by the explicit
symmetry breaking term $K_U$.
Thus the fluctuations of the diagonal modes are gapped.
Alternatively, in the case where the symmetry breaking of the
diagonal excitons is spontaneous and $U=0$, there will
be $N$ Goldstone modes.
In the following we will focus on the dynamics of 
the remaining $N^2 - N$ off-diagonal modes
which will be the same in both cases.

The effective action for the off-diagonal modes can be readily
obtained from the full action (\ref{theaction}).
The phase locking of the diagonal modes 
plays two important roles here.
First, it create mass for some of the off-diagonal modes.
We pick out terms in Eq. (\ref{theaction}) which 
are proportional to $\chi_o^2$ and 
products of the the diagonal elements $\bar \chi^{aa}$.
The derivation is straightforward and only the results are shown here
(for details see Appendix B).
This is analogous to the Higgs mechanism.
Depending on the sign of the `mass' term for the off-diagonal modes,
\bq
\tilde K_2 \sum_{a \neq b} \chi^{ab} \chi^{ba}
= \tilde K_2 \chi_o^2 \sum_{a \neq b} 
\cos( \theta^{ab} + \theta^{ba} )
\label{mass}
\eq
with
$\tilde K_2 \equiv 2 \chi_d^{2} \left[ K_4 + N K_6^{'} \left( \chi_d^2 + (N-1) \chi_o^2 \right) + 3 K_6 \chi_d^2 \right]$,
$\chi$ is constrained to be either Hermitian or anti-Hermitian at low energy.
Second, the coherent diagonal modes make the dynamics of the
off diagonal mode relativistic.
Once $\chi$ is constrained to be Hermitian or anti-Hermitian matrix,
the excitons $\chi^{ab}$  and $\chi^{ba}$ are no longer independent,
but they are anti-particles to each other.
The physical picture is that two excitons 
$\chi^{ab}$ and $\chi^{ba}$
scatter with each other to become two diagonal excitons 
$\chi^{aa}$ and $\chi^{bb}$.
The diagonal excitons are condensed into vacuum and
from the point of view of the off diagonal excitons, 
this is an annihilation process of two excitons.
Conversely, two off diagonal excitons can be created out of vacuum. 
Since the low energy theory is relativistic, 
the effective action has symmetric form in space and time directions.
It reduces to a Hermitian matrix model
(for details see Appendix B),
\bqa
S  & = &
- \frac{K}{4} \sum_{<i,j>} tr^{'} ( \chi_i^\dagger \chi_j + h.c.) 
 + \sum_{n \geq 3} \tilde K_n \sum_i tr^{'} \chi_i^n  \nn
\label{H}
\eqa
for $\tilde K_2  <  0$
or an anti-Hermitian matrix model,
\bqa
S  & = &
- \frac{K}{4} \sum_{<i,j>} tr^{'} ( \chi_i^\dagger \chi_j + h.c.) 
 + \sum_{n \geq 2} \tilde K_{2n} \sum_i tr^{'} \chi_i^{2n}  \nn
\label{antiH}
\eqa
for $\tilde K_2  >  0$.
Here $i,j$ are indices for lattice in D-dimensional Euclidean space-time.
$\chi$ is Hermitian (anti-Hermitian) matrix without diagonal element in (\ref{H}) ((\ref{antiH})).
The $'$ sign in the trace is to recall that there is no diagonal element in $\chi$. 
The explicit form of $\tilde K_n$ is also obtained in the Appendix B.
The first term in the above actions is the kinetic energy term for the exciton condensate
and the second term, the potential energy term.

Some remarks are in order for these models.
First, (\ref{H}) and (\ref{antiH}) are two different theories in general.
However they are equivalent in some special cases.
The Hermitian matrix model with $\tilde K_{2n-1} = 0$ and 
$\tilde K_{2n} = k_{2n}$ is equivalent to the anti-Hermitian matrix model 
with $\tilde K_{2n} = (-1)^n k_{2n}$.
They are related to each other by the transformation $\chi^{'} = i \chi$.
Similar equivalence also exist within the Hermitian matrix model.
The Hermitian matrix model with $\tilde K_{n}$ is equivalent 
to the Hermitian matrix model with $\tilde K_{n}^{'} = (-1)^n \tilde K_{n}$.
They are related by the transformation $\chi^{'} = -\chi$. 
Second, the lowest order nonzero interaction is most important 
in the second terms of (\ref{H}) and (\ref{antiH})
if $\chi_o << 1$.
This is because the effective phase stiffness for the off-diagonal modes is suppressed by factor of $\chi_o^n$.
More precisely, nonzero $\tilde K_l$ with smallest $l$ determines the dynamics of the theories
if $|\tilde K_l| >> \chi_o^m |\tilde K_{l+m}|$ for all $m$.
The above consideration greatly reduces the number of theories to consider.
Up to the quartic order, we have only the following three different low energy theories, 
\begin{itemize}
\item[A.] Hermitian matrix model with $K_3 \neq 0$
\item[B.] anti-Hermitian matrix model with $K_4 < 0$ 
(equivalently, Hermitian matrix model with $K_3 = 0$ and $K_4 < 0$)
\item[C.] anti-Hermitian matrix model with $K_4 > 0$ 
(equivalently, Hermitian matrix model with $K_3 = 0$ and $K_4 > 0$)
\bq
\label{category}
\eq
\end{itemize}
In this paper we examine the low energy physics of the above three cases in detail.
We note that the cases in parenthesis in cases B and C are non-generic
and require some tuning parameter to set $\tilde K_3 = 0$.
We do not consider non-generic cases with $K_3 = K_4 = 0$ where 
more low energy degrees of freedom may emerge.

\section{Emergent U(1) gauge theory in the matrix models}

In this section we will analyze the low energy physics of the three theories (\ref{category}) in the strong coupling limit $|\tilde K_n| \chi_o^n >> 1$.
The degrees of freedom of the (anti) Hermitian matrix will 
be further reduced by the dynamical constraints imposed by the strong interaction.
We will see that the low energy degrees of freedom  
are U(1) gauge field and particles which
carry only fractional quantum number of exciton.
The statistics and spin of the fractionalized particle will be determined
by the coupling constants.

\subsection{Emergence of fractionalized boson and U(1) gauge field}

Here we consider the first case in (\ref{category}) : the Hermitian matrix model (\ref{H}) with $\tilde K_3 \neq 0$.
Without loss of generality we assume $\tilde K_3 < 0$.
For simplicity we consider the case with $\tilde K_n < 0$ for all $n$ even though
only the $\tilde K_3$ is important for small $\chi_o$.
The $\tilde K_3$ term in (\ref{H}), 
\bqa
\tilde K_3 \chi_o^3 \sum^{'}_{a,b,c} \cos ( \theta^{ab} + \theta^{bc} + \theta^{ca} )
\label{b1_K3}
\eqa
leads to constraints (in addition to the Hermitivity),
\bqa
\theta^{ab} + \theta^{bc} + \theta^{ca}  & = & 0 
\label{c3}
\eqa
in the strong coupling limit $|\tilde K_3| \chi_o^3  >> 1$.
These conditions also constrain sum of $n \geq 4$ phases.
For example, by summing the two conditions
\bqa
\theta^{ab} + \theta^{bc} + \theta^{ca}  & = & 0, \nn
\theta^{ac} + \theta^{cd} + \theta^{da}  & = & 0
\eqa
and using the Hermitivity $\theta^{ca} + \theta^{ac}  = 0$,
we obtain
\bqa
\theta^{ab} + \theta^{bc} + \theta^{cd} + \theta^{da}  & = & 0.
\eqa
Thus the higher order interaction terms  in (\ref{H}) are also minimized with $\tilde K_n <0$.
The number of independent constraints imposed by (\ref{c3}) is $\frac{(N-1)(N-2)}{2}$\cite{COUNTING}.
leading to 
$\frac{N(N-1)}{2} - \frac{(N-1)(N-2)}{2} = N-1$
soft modes. 
These modes can be parametrized by the $N$ phases $\phi^a$ so that
\bq
\theta^{ab}_i = \phi^a_i - \phi^b_i
\label{boson_decompose}
\eq
with a redundancy $\phi^a_i \sim \phi^a_i + \phi_i$.
The (N-1) low energy bosons are the soft modes 
associated with the $U(1)^{N-1}$ symmetry
of the Landau theory (\ref{theaction}).
Note that $e^{i \phi^a_i}$ carries only 
one flavor quantum number while exciton carries two.
Thus we refer to this boson as {\it slave} boson.
At this stage, the slave boson is just a book-keeping tool.
It may or may not appear as low energy excitation. 
If the slave boson appears as true low energy excitation,
we will refer to it as {\it fractionalized} boson, the term 
which is mainly used in literatures.

\begin{figure}
        \includegraphics[height=7cm,width=7cm]{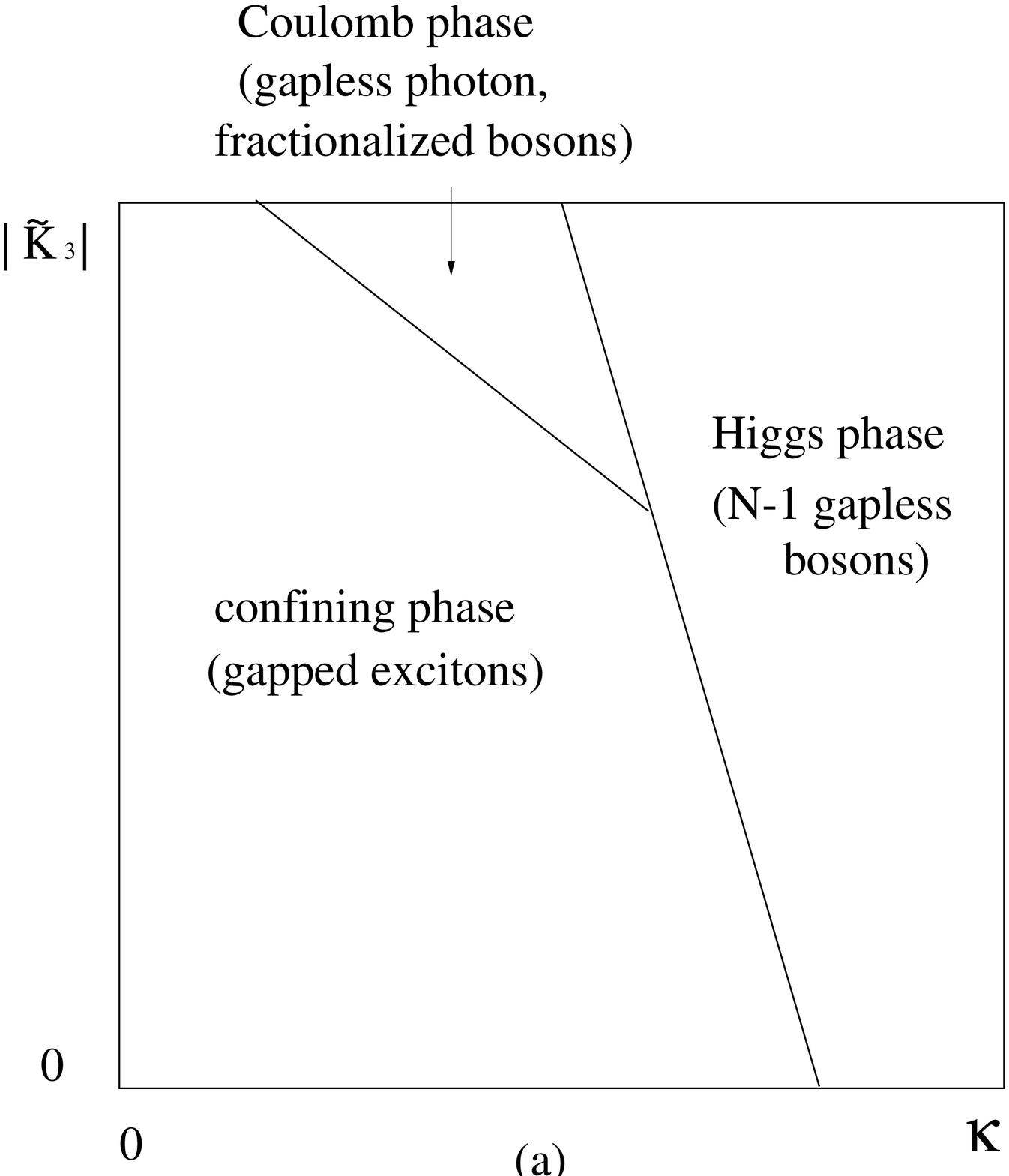}
        \includegraphics[height=7cm,width=7cm]{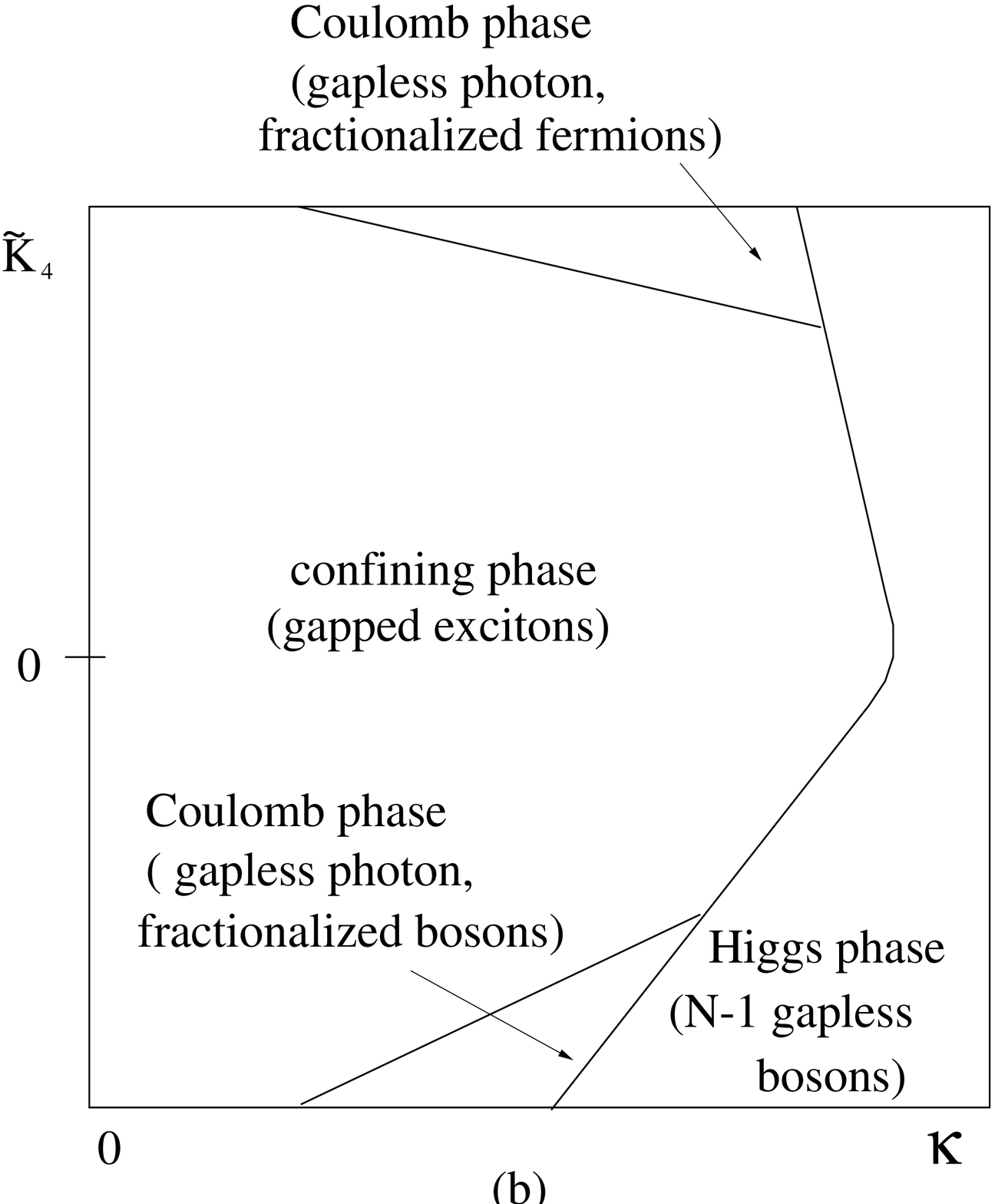}
\caption{
Schematic phase diagrams of
(a) the Hermitian matrix model with  $\tilde K_3 \neq 0$
and (b) the anti-Hermitian matrix model
(equivalently the Hermitian matrix model with $\tilde K_3 = 0$ )
in $d=3$.
The Ising transition line is not shown in (b).
}
\label{fig:phase}
\end{figure}

For the low energy modes $\phi_i^a$ there is no potential energy
since Eq. (\ref{b1_K3}) is minimized.
The effective action becomes
\bqa
S & = & -\frac{\kappa}{4} \sum_{a<b} \sum_{<i,j>} \left[ e^{ i ( \phi^{a}_i - \phi^{a}_j )} e^{ - i ( \phi^{b}_i - \phi^{b}_j )} + c.c. \right],
\eqa
where $\kappa = 2 K \chi_o^2$.
Introducing the Hubbard-Stratonovich transformation for the quartic term,
\bqa
&& \exp\Big( \frac{\kappa}{4} 
e^{ i ( \phi^{a}_i - \phi^{a}_j  ) }  
e^{ i ( \phi^{b}_j - \phi^{b}_i  ) }  
+ c.c. \Bigr) \nn
& = &
\left( \frac{ \kappa}{4\pi} \right)^2
\int
d\eta_{ij}^{ab}
d\eta_{ij}^{ab*}
d\eta_{ji}^{ab}
d\eta_{ji}^{ab*} \nn
&& \times
\exp \Bigl(  - \frac{\kappa}{4} \Bigl[
| \eta_{ij}^{ab} |^2 + |\eta_{ji}^{ab}|^2 \nn
&&   -  \eta_{ij}^{ab} e^{ i ( \phi^{a}_i - \phi^{a}_j ) }
   -  \eta_{ij}^{ab*} e^{ i ( \phi^{b}_j - \phi^{b}_i ) } \nn
&& -  \eta_{ji}^{ab} e^{ -i ( \phi^{a}_i - \phi^{a}_j ) }
   -  \eta_{ji}^{ab*} e^{ -i ( \phi^{b}_j - \phi^{b}_i ) }
\Bigr] \Bigr),
\eqa
and parameterizing the two complex Hubbard-Stratonovich fields by four real variables,
\bqa
\eta_{ij}^{ab} & = & |\tilde \eta_{ij}^{ab}| e^{  w^{ab}_{ij} + i ( a^{+ab}_{ij} - a^{ab}_{ij} ) }, \nn
\eta_{ji}^{ab} & = & |\tilde \eta_{ij}^{ab}| e^{ -w^{ab}_{ij} + i ( a^{+ab}_{ij} + a^{ab}_{ij} ) },
\eqa
we obtain the action
\bqa
S_{b} & = &   
\frac{\kappa}{4} \sum_{<i,j>,a<b} \Bigl[
2 |\tilde \eta_{ij}^{ab}|^2 \cosh (  2  w^{ab}_{ij} ) \nn
&&
- |\tilde \eta_{ij}^{ab}|  e^{w_{ij}^{ab}  +i  ( a^{+ab}_{ij} - a^{ab}_{ij} )} e^{ i ( \phi^{a}_i - \phi^{a}_j ) } \nn
&& - |\tilde \eta_{ij}^{ab}|  e^{-w_{ij}^{ab} +i  ( a^{+ab}_{ij} + a^{ab}_{ij} )} e^{ -i ( \phi^{a}_i - \phi^{a}_j ) } \nn
&&
- |\tilde \eta_{ij}^{ab}|  e^{w_{ij}^{ab} +i  ( - a^{+ab}_{ij} + a^{ab}_{ij} )} e^{ i ( \phi^{b}_j - \phi^{b}_i ) } \nn
&& - |\tilde \eta_{ij}^{ab}|  e^{-w_{ij}^{ab} +i  ( - a^{+ab}_{ij} - a^{ab}_{ij} )} e^{ -i ( \phi^{b}_j - \phi^{b}_i ) }
\Bigr] .
\label{b1_s}
\eqa
Among the auxiliary fields, there is only one massless modes associated with the U(1) gauge symmetry,
\bqa
\phi^{a }_i & \rightarrow & \phi^{a}_i + \varphi_i, \nn
a^{ab}_{ij} & \rightarrow & a^{ab}_{ij} + \varphi_i - \varphi_j.
\eqa
We retain the U(1) gauge field and neglect the fluctuations of other auxiliary fields.
In general the action (\ref{b1_s}) is not real but
reality is restored at the saddle point where $a^{+ab}_{ij}$ and $w_{ij}^{ab}$ have
saddle points in the imaginary direction\cite{LEE}.
In our case, by comparing the first and last terms in Eq. (\ref{b1_s}),
we note that the only solution where the effective hopping is 
the same for all species of bosons is
\bq
a^{+ab'}_{ij} = w_{ij}^{ab} =0.
\eq
The massless mode is one where $a_{ij}^{ab}$ is independent of 
the flavor index, i.e., $a^{ab}_{ij} = a_{ij}$ and
the low energy action becomes
\bqa
S_{b} & = &   
 \frac{\kappa}{4} (N-1) \sum_a \sum_{<i,j>} \Bigl[
  \eta^2 - \eta  e^{ i ( \phi^{a}_i - \phi^{a}_j - a_{ij}) }
- c.c.  \Bigr], \nn
\label{boson_action}
\eqa
where $|\tilde \eta_{ij}^{ab}| = \eta$.
This is the familiar problem of compact U(1) gauge theory coupled with N bosons.
The bare gauge coupling is infinite.
At low energy the gauge coupling is renormalized to be finite.
For example, by expanding in $\kappa$ and integrating out bosons which hop 
around a plaquette, we produce a gauge coupling associated with the Maxwellian term 
on the unit plaquette in the D-dimensional (hyper) cubic lattice, 
$g^2 \sim \frac{1}{t^4 N}$ with $t \equiv \frac{\kappa}{4}(N-1) \eta$.
In the small $g$ limit the confinement phase occurs.
The slave bosons do not appear as excitation.
All excitations are gapped.
This is the disordered phase of the off diagonal excitons.
In the large $t$ limit the bosons have phase coherence (Higgs phase).
One of the N boson is eaten by the massive U(1) gauge field.
There are remaining $N-1$ massless modes.
In terms of original exciton, 
this is the phase where the $N^2 - N$ off diagonal excitons are bose condensed. 
Due to locking with the already correlated diagonal excitons,
only $N-1$ modes survive as the Goldstone modes associated with the
off diagonal exciton condensates.
This is the expected result from the $U(1)^{N-1}$ symmetry as discussed in Sec. II.
These are the possible phases in 2+1D.
In 2+1D the deconfinement phase may exist only at the critical point
between the confining phase and the Higgs phase.
In higher dimension the Coulomb phase can occur between the confinement and Higgs phases.
In the Coulomb phase the photon emerges as a massless collective excitation of excitons. 
The physical meaning of the photon in terms of original exciton will be discussed in the Sec. V. A. 
In this phase the bosons are gapped and interact by exchange of photons.
The schematic phase diagram in 3+1D is shown in Fig. \ref{fig:phase} (a).

Since the fractionalized boson emerges from the exciton which is charge neutral
to electromagnetic field, it is obvious that it is charge neutral as well.
The fractionalized boson is structureless and it seems obvious that the boson has spin $0$.
By spin $0$ we mean that the boson transforms trivially under Lorentz transformation
in the continuum limit.
However we need to take into account the coupling to the gauge field
in order to determine the spin of the boson at low energy.
Let us follow only one species of boson by integrating out the rest of the bosons.
For small $t$ we obtain
\bqa
S & = & 
- t \sum_{<i,j>} \left[ 
 e^{ i ( \phi_i - \phi_j - a_{ij} )} + c.c.  \right] \nn
&& - \sum_{C} \frac{1}{2 g_C^2} \cos b_C.
\label{b1_u1}
\eqa
Here the flavor index for the remaining boson is dropped.
$C$ denotes loop in space-time traced by the integrated bosons.
$b_C \equiv \oint_C a$ is the flux associated with the loop.
$g_C^2 \sim \frac{1}{(N-1) t^{L(C)}}$ is the generalized gauge coupling 
for the non-local gauge action with
$L(C)$, the length of the loop $C$.
In the massive phase for the fractionalized boson with $t << 1$
small loop contributions dominate the gauge action.
Note that the gauge coupling can be made small with large N even for small $t$.
In the Coulomb phase
the gauge field has the saddle point value $a_{ij} = 0$. 
Hence the background gauge field for the remaining boson is smoothly varying 
in space-time.
Thus the boson remains to be structureless in the long-wavelength limit.
The continuum Lagrangian for the emergent boson becomes
\bqa
{\cal L} & = & 
\sum_\mu \tilde t_\mu (  \partial_\mu  \phi -  a_\mu )^2  
+ \frac{1}{2 \tilde g^2} \sum_{\mu,\nu} f_{\mu \nu}^2,
\eqa
where  $\tilde t_\mu$ and $\tilde g$ are the renormalized
phase stiffness and gauge coupling respectively.
$f_{\mu \nu} = \partial_\mu a_\nu - \partial_\nu a_\mu$
is the field strength tensor for the emergent non-compact gauge field.
The fractionalized boson is in the disordered phase due to
the proliferation of vortices and emerges
as gapped excitation which has
the relativistic dispersion,
\bqa
\omega_k = \sqrt{ v^2 k^2 + m^2},
\eqa
where $v = \sqrt{t_1/t_0}$ is the velocity 
with $t_0 = \tilde t_{\mu=0}$, 
$t_1 = \tilde t_{\mu \neq 0}$
and $m$, the mass gap.
The mass gap vanishes at the transition to the Higgs phase.
If the background gauge field had turned out to be rapidly varied in the lattice scale, 
the emergent particle in the long-wavelength limit
would have been different from the boson in the lattice scale.
If this were the case, multiple components which correspond to
spin quantum numbers could emerge in the continuum limit
from the doubling of unit cell.
This turns out to be the case for emergent fermions,
as we shall see later.

\subsection{Emergence of fractionalized boson, Ising mode and U(1) gauge field}

In this section we consider the second case in (\ref{category}).
For simplicity, we consider the Hermitian matrix model with $\tilde K_{2n} < 0$  for all $n$.
Equivalently, we may use the anti-Hermitian matrix model with $\tilde K_{2n}  = (-1)^{n-1} |\tilde K_{2n} |$.
Here we use the Hermitian matrix model because it is more convenient to make the coupling constant have same sign.
The fourth order term in (\ref{antiH}), 
\bqa
\tilde K_4  \chi_o^4 \sum^{'}_{a,b,c,d} \cos ( \theta^{ab} + \theta^{bc} + \theta^{cd} + \theta^{da} )
\eqa
leads to the constraints
\bqa
\theta^{ab}_i + \theta^{bc}_i + \theta^{cd}_i + \theta^{da}_i & = & 0
\label{c4}
\eqa
for $|\tilde K_4 | \chi_o^4 >> 1$.
Surprisingly, these conditions also constrain the sum of three exciton phases.
By summing the three equations,
\bqa
\theta_i^{ab} + \theta_i^{bd} + \theta_i^{dc} + \theta_i^{ca} & = & 0, \nn
\theta_i^{ac} + \theta_i^{cb} + \theta_i^{bd} + \theta_i^{da} & = & 0, \nn
\theta_i^{ad} + \theta_i^{dc} + \theta_i^{cb} + \theta_i^{ba} & = & 0, 
\eqa
and using the Hermitivity, $\theta_i^{ab} + \theta_i^{ba} = 0$,
we obtain
\bqa
2 ( \theta_i^{cb} + \theta_i^{bd} + \theta_i^{dc} )  & = & 0.
\eqa
Therefore the sum of three phases are fixed up to $Z_2$ degree of freedom,
\bqa
\theta_i^{cb} + \theta_i^{bd} + \theta_i^{dc} & = & 0~ \mbox{~or}~ \pi.
\eqa
The constraint for the four phases (\ref{c4}) is less restrictive than 
the constraint for the three phases (\ref{c3}) in the previous section
and an extra $Z_2$ degrees of freedom is needed.
Thus the low energy degrees of freedom are parameterized by the
$N$ bosonic variables and a $Z_2$ variable, 
\bq
\theta^{ab}_i = \phi^a_i - \phi^b_i + \frac{\pi}{2} (1 + z_i),
\label{boson_decompose2}
\eq
with $z_i = \pm 1$.
The Hermitivity is satisfied with the decomposition.
The $\tilde K_{2n}$ term with $n \geq 3$ is also minimized 
because
\bq
\cos \left(
\theta_i^{a_1 a_2} + 
\theta_i^{a_2 a_3} + ...
\theta_i^{a_{2n} a_1} 
\right) = 1.
\eq
Thus one exciton field is decomposed into three degrees of freedom.
The two U(1) phases are the same modes which appear in the previous section.
The Ising mode is the flavor independent $Z_2$ degree of freedom.
It is noted that $z_i$ is an independent degree of freedom
because it can not be absorbed into $\phi^a$ or $\phi^b$.
This can be illustrated by the relation
$z_i = -e^{i( \theta^{ab} + \theta^{bc} + \theta^{ca} )}$ whose value is
not affected by the change of $\phi^a$. 
The $z_i$ describes the low energy $Z_2$ degree of freedom for a composite of three excitons.
Another important consequence of this relation
is that because $z_i$ is fully determined in terms of $\theta^{ab}$'s,
there is no gauge symmetry under which the $Z_2$ variable transforms nontrivially.
This is contrary to the gauge symmetry of the U(1) variables, $\phi^a_i \rightarrow \phi^a_i + \phi_i$.
The structure of local symmetry in the decomposition determines how the slave particles are coupled to emergent gauge field at low energy.
There will be U(1) gauge field coupled to the U(1) fields, but
no gauge coupling to the Ising mode.

The effective action for the low energy modes becomes
\bqa
S & = & -\frac{\kappa}{4} \sum_{a<b} \sum_{<i,j>} 
\left[ 
e^{ i ( \phi^{a}_i - \phi^{a}_j )} 
e^{ - i ( \phi^{b}_i - \phi^{b}_j )} 
e^{i \frac{\pi}{2} (z_i - z_j)} + c.c. 
\right], \nn
\eqa
where $\kappa = 2 K \chi_o^2$.
We first decompose the U(1) variables from the $Z_2$ variables,
\bqa
&& \exp\Big( \frac{\kappa}{4} 
e^{ i ( \phi^{a}_i - \phi^{a}_j ) - i ( \phi^{b}_i - \phi^{b}_j )} e^{i \frac{\pi}{2} (z_i - z_j)} 
+ c.c. \Bigr) \nn
& = &
\left( \frac{ \kappa}{4\pi} \right)^2
\int
d\xi_{ij}^{ab}
d\xi_{ij}^{ab*}
d\xi_{ji}^{ab}
d\xi_{ji}^{ab*} \nn
&& \times
\exp \Bigl(  - \frac{\kappa}{4} \Bigl[
| \xi_{ij}^{ab} |^2 + |\xi_{ji}^{ab}|^2 \nn
&&   -  \xi_{ij}^{ab} e^{ i  ( \phi^{a}_i - \phi^{a}_j ) - i ( \phi^{b}_i - \phi^{b}_j )  }
   -  \xi_{ij}^{ab*}  e^{i \frac{\pi}{2} (z_i - z_j)} \nn
&&   -  \xi_{ji}^{ab} e^{ -i  ( \phi^{a}_i - \phi^{a}_j ) + i ( \phi^{b}_i - \phi^{b}_j )  }
   -  \xi_{ji}^{ab*}  e^{-i \frac{\pi}{2} (z_i - z_j)} 
\Bigr] \Bigr).
\eqa
With the parameterization for the Hubbard-Stratonovich fields
\bqa
\xi_{ij}^{ab} & = & |\tilde \xi_{ij}^{ab}| e^{  x^{ab}_{ij} + i ( c^{+ab}_{ij} + c^{ab}_{ij} ) }, \nn
\xi_{ji}^{ab} & = & |\tilde \xi_{ij}^{ab}| e^{ -x^{ab}_{ij} + i ( c^{+ab}_{ij} - c^{ab}_{ij} ) },
\eqa
the action becomes
\bqa
S & = &   
\frac{\kappa}{4} \sum_{<i,j>,a<b} \Bigl[
2 |\tilde \xi_{ij}^{ab}|^2 \cosh ( 2 x^{ab}_{ij} ) \nn
&&
- |\tilde \xi_{ij}^{ab}|  e^{x_{ij}^{ab}  +i  ( c^{+ab}_{ij} + c^{ab}_{ij} )} 
	e^{ i ( \phi^{a}_i - \phi^{a}_j ) - i ( \phi^{b}_i - \phi^{b}_j ) } \nn
&& - |\tilde \xi_{ij}^{ab}|  e^{-x_{ij}^{ab} +i  ( c^{+ab}_{ij} - c^{ab}_{ij} )} 
	e^{ -i ( \phi^{a}_i - \phi^{a}_j ) +  i ( \phi^{b}_i - \phi^{b}_j )} \nn
&&
- |\tilde \xi_{ij}^{ab}|  e^{x_{ij}^{ab} +i  ( - c^{+ab}_{ij} - c^{ab}_{ij} )} 
e^{i \frac{\pi}{2} (z_i - z_j)} \nn
&& - |\tilde \xi_{ij}^{ab}|  e^{-x_{ij}^{ab} +i  ( - c^{+ab}_{ij} + c^{ab}_{ij} )}  
e^{-i \frac{\pi}{2} (z_i - z_j)}
\Bigr] .
\eqa
The action is not Hermitian.
However the Hermitivity is restored at the saddle point 
$x_{ij}^{ab} = i \tilde x_{ij}^{ab}$ and $c^{+ab}_{ij} =  i \tilde c_{ij}^{+ab}$
with real $\tilde x_{ij}^{ab}$ and $ \tilde c_{ij}^{+ab}$.
At the saddle point we have real action, 
\bqa
S & = &   
\frac{\kappa}{4} \sum_{<i,j>,a<b} \Bigl[
\xi_z \xi_\phi
- \xi_z e^{ i ( \phi^{a}_i - \phi^{a}_j ) - i ( \phi^{b}_i - \phi^{b}_j ) } \nn
&& - \xi_\phi e^{i \frac{\pi}{2} (z_i - z_j)}
\Bigr] + c.c.,
\eqa
where
$\xi_z \equiv |\tilde \xi_{ij}^{ab}|  e^{- \tilde c^{+ab}_{ij}  + i ( \tilde x_{ij}^{ab} +c^{ab}_{ij} )}$ and
$\xi_\phi \equiv |\tilde \xi_{ij}^{ab}|  e^{\tilde c^{+ab}_{ij}  + i ( \tilde x_{ij}^{ab} -c^{ab}_{ij} )}$.
All the fluctuations of the auxiliary field around the saddle point is massive.
Thus the U(1) phase modes and the Ising mode are decoupled at low energy.
The saddle point values of the auxiliary fields are determined from the self-consistency conditions,
\bqa
\xi_z & = & \left<  e^{i \frac{\pi}{2} (z_i - z_j)} \right>, \nn
\xi_\phi & = & \left< e^{ i ( \phi^{a}_i - \phi^{a}_j ) - i ( \phi^{b}_i - \phi^{b}_j ) } \right>.
\eqa
The solution with $\xi_z, \xi_\phi > 0$ satisfies the self-consistency.
That is, positive $\xi_\phi$ gives $\left<  e^{i \frac{\pi}{2} (z_i - z_j)} \right> > 0$
and positive $\xi_z$, in turn, gives $\left< e^{ i ( \phi^{a}_i - \phi^{a}_j ) - i ( \phi^{b}_i - \phi^{b}_j ) } \right> > 0$.
Thus the $Z_2$ variable is ferromagnetically coupled with each other.
The final result is the same as that of the usual mean-field decomposition with $AB \rightarrow <A>B + A<B> - <A><B>$.
The merit of the above procedure is that one can easily identify the low energy degrees of freedom beyond
the saddle point approximation\cite{LEE}.

The quartic term of the U(1) phase modes are decoupled 
by the same transformation used in the previous section.
Finally, we obtain the compact U(1) gauge theory coupled with $N$ bosons
and the ferromagnetically coupled Ising field,
\bqa
S & = & 
-t \sum_a \sum_{<i,j>} \Bigl[ e^{ i ( \phi^{a}_i - \phi^{a}_j - a_{ij} ) } + c.c.  \Bigr] \nn
&& - t_z \sum_{<i,j>} z_i z_j,
\label{b2_s}
\eqa
where $t = \frac{\kappa}{4} (N-1) \xi_z \eta$ 
and
$t_z = \frac{\kappa}{4} N (N-1) \xi_\phi$.

The physics of the U(1) gauge sector is the same as the one discussed in the previous section.
On the other hand, the Ising variable undergoes a second order phase transition from the disordered phase to ordered phase as $t_z$ increases. 
The low energy dynamics of the Ising mode 
is described by the real Klein-Gordon model.
The energy dispersion becomes
\bqa
\omega_k & = & \sqrt{ v_I^2 k^2 + m_I^2},
\eqa
where $v_I$ is the velocity
and $m_I$, the mass gap.
The mass gap vanishes only at the critical point.
In the ordered phase the composite field of three excitons has long range correlation.
Since the low energy theories are decoupled for the U(1) and the Ising sectors,
the Ising transition can occur within any phase in the phase diagram of Fig. \ref{fig:phase} (b).
If the Ising transition occurs within the Higgs phase,
the Higgs phase of the fractionalized boson without long range correlation
of exciton is possible.
In this phase pairs of off diagonal excitons are condensed, i.e.,
$\left< e^{i (\theta^{ab} + \theta^{cd}) } \right> \neq 0$.
This is in contrast to case A.

\subsection{Emergence of fractionalized fermion, Ising mode and U(1) gauge field}

In this section we consider the third case in (\ref{category}).
Here we choose to use the anti-Hermitian matrix model with $\tilde K_{2n} > 0$ which 
is equivalent to the Hermitian matrix model with $\tilde K_{2n} = (-1)^n |\tilde K_{2n}|$.
The potential energy
\bq
K_{2n} tr^{'} \chi^{2n} = K_{2n} \chi_o^{2n} \sum_{a_1,..,a_{2n}} \cos( \theta^{a_1 a_2} +  \theta^{a_2 a_3} +... + \theta^{a_{2n} a_1} )
\label{f1p}
\eq
is minimized when
\bqa
\theta^{a_1 a_2}_i + \theta^{a_2 a_3}_i + ...  + \theta^{a_{2n} a_1}_i 
& = & \pi.
\label{fn}
\eqa
These become constraints if $K_{2n} \chi_o^{2n} >> 1$.
The condition (\ref{fn}) defines the `low energy valley' in the phase space of $\theta^{ab}$.
However there is no bosonic parameterization for the low energy modes
because of frustration in the interaction.
This can be argued in the following ways.
Firstly, (\ref{fn}) is not consistent with the anti-Hermitian matrix. 
The constraint 
$\theta^{ab}_i+ \theta^{bc}_i + \theta^{cd}_i + \theta^{da}_i    =  \pi$
leads to 
$\theta^{ab}_i+ \theta^{ba}_i + \theta^{bc}_i + \theta^{cb}_i    =  \pi$
for $d = b$.
However this is inconsistent with the anti-Hermitivity condition
$\theta^{ab}_i+ \theta^{ba}_i =  \pi$ which leads to
$\theta^{ab}_i+ \theta^{ba}_i + \theta^{bc}_i + \theta^{cb}_i    =  0$ mode $2 \pi$.
Secondly, the conditions (\ref{fn}) are frustrated among themselves.
Summing the two equations 
$\theta^{ab}_i+ \theta^{bc}_i + \theta^{cd}_i + \theta^{da}_i    =  \pi$, 
$\theta^{ae}_i + \theta^{ec}_i + \theta^{cb}_i + \theta^{ba}_i   =   \pi$
and using the anti-Hermitivity
$ \theta^{ab}_i + \theta^{ba}_i = \pi$,
$ \theta^{bc}_i + \theta^{cb}_i = \pi$
we obtain
$\theta^{cd}_i + \theta^{da}_i + \theta^{ae}_i + \theta^{ec}_i   =  0$ mod $2\pi$.
In other words, the set of constraint given by 
Eq. (\ref{fn}) is inherently frustrated and can not
be all satisfied by any set of $\{ \theta^{ab} \}$ where
$\theta^{ab}$ are c-numbers.

We are thus motivated to introduce anti-commutativity,
\bqa
e^{i \theta^{ab}_i} & = & i \psi^{a *}_i \psi^b_i. 
\label{fermion_decompose1}
\eqa
Here $\psi$ is Grassmann fields.
The $i$ is introduced to satisfy anti-Hermitivity, 
$e^{i \theta^{ba}_i} = - \left( e^{i \theta^{ab}_i}  \right)^*$.
Because of the anti-commutativity, 
all of the constraints can be satisfied, i.e.,
\bqa
\left< e^{i (\theta^{ab}_i+ \theta^{bc}_i + \theta^{cd}_i + \theta^{da}_i) } \right>
& = & \left< 
\psi^{a *}_i \psi^b_i
\psi^{b *}_i \psi^c_i
\psi^{c *}_i \psi^d_i
\psi^{d *}_i \psi^a_i
\right> \nn
& = & - \left< 
\psi^a_i \psi^{a *}_i 
\psi^b_i \psi^{b *}_i 
\psi^c_i \psi^{c *}_i 
\psi^d_i \psi^{d *}_i 
\right>  \nn
& = &-1.
\eqa
Here the fermionic functional integral is performed with the measure,
\bqa
\int \Pi_{a<b} \left[ \frac{d \theta^{ab}}{2\pi} \right] & = & 
\int \Pi_a \left[ d \psi^{a*} d \psi^a \right] {\cal M}, 
\label{f1_int}
\\
{\cal M} & = &      
i^N 
e^{- i  \sum_a \psi^{a} \psi^{a*} },
\label{f1_measure}
\eqa
where the integration in the left hand side of Eq. (\ref{f1_int}) 
is understood to be performed over the low energy valley 
in the energy landscape for the exciton phases.
However the decomposition (\ref{fermion_decompose1}) is not complete yet.
This is because the fermionic decomposition also predicts 
$\left< e^{ i(\theta^{ab}_i + \theta^{bc}_i + \theta^{ca}_i )} \right> \neq 0$,
while it is zero  in the original exciton representation.
It should be zero due to the $Z_2$ symmetry ($\theta^{ab}_i \rightarrow \theta^{ab}_i + \pi$) 
of the action (\ref{antiH}).
Thus we include the extra $Z_2$ degree of freedom,
\bqa
e^{i \theta^{ab}_i} & = & i \psi^{a *}_i \psi^b_i e^{i \frac{\pi}{2} (1 + z_i)},
\label{fermion_decompose}
\eqa
with $z_i = \pm 1$.
The situation is similar to the bosonic case in the previous section.

With the decomposition we have the consistency for elementary integrations,
\bqa
&& \left< 1 \right> = \frac{1}{2} \sum_z \int  \Pi_a \left[ d \psi^{a*} d \psi^a \right] {\cal M} = 1, \nn
&& \left< e^{i \theta^{bc}} \right> = 
\frac{1}{2} \sum_z \int \Pi_a \left[ d \psi^{a*} d \psi^a \right] {\cal M} i z \psi^{b *} \psi^c = 0, \nn
&& \left< 
e^{i \theta^{bc}} 
e^{-i \theta^{bc}} 
\right> = 
\frac{1}{2} \sum_z \int \Pi_a \left[ d \psi^{a*} d \psi^a \right]  {\cal M} 
 \psi^{b *} \psi^c 
 \psi^{c *} \psi^b  
= 1, \nn
\label{f1_integ}
\eqa
and the potential energy is minimized, 
\bqa
\left< 
\cos( \theta^{a_1 a_2}_i + \theta^{a_2 a_3}_i + ...  + \theta^{a_{2n} a_1}_i )
\right> & = & -1
\label{f1_ex}
\eqa
for all $\{ a_1,a_2,...,a_{2n} \}$.
At the same time the product of odd number of exciton fields has zero expectation value
because of the fluctuating $Z_2$ degree of freedom.  
This is also consistent with the property of original field.
Thus these fermionic variables and the $Z_2$ spins are the low energy modes of the system 
in the strong coupling regime.
Note that the fermionic decomposition is over-estimating the 
expectation value of potential energy because
the absolute value of the left hand side 
in Eq. (\ref{f1_ex}) should be less than $1$.
For a quantitative agreement one may have to introduce
four or higher fermion terms in the measure.
Here we proceed with the measure (\ref{f1_measure}) because
the four or higher fermion interaction is irrelevant at low energy
for $d \geq 2$.

The partition function is written as
\bqa
Z & = & \int \sum_z \Pi_{i,a} \left(
i 
d \psi_i^{a*} 
d \psi_i^{a} 
\right)
e^{-S},
\label{f1z}
\eqa
where 
\bqa
S & = &  i \sum_i \sum_a \psi_i^{a} \psi_i^{a*} \nn
&& + \frac{\kappa}{4} \sum_{a<b} \sum_{<i,j>} 
\left[  \psi^{a *}_i \psi^{a }_j \psi^{b*}_j \psi^{b}_i   e^{i \frac{\pi}{2} (z_i - z_j)}
+ c.c. \right]
\label{f1s}
\eqa
with $\kappa = 2 K \chi_o^2$.
The measure of the integration can be cast into the conventional form by the transformation\cite{GRASSMANN}.
\bqa                
\psi_i^a & = &  e^{i \frac{\pi}{4} } \psi_i^{a'}, \nn
\psi_i^{a*} & = & e^{i \frac{\pi}{4} } \psi_i^{a'*}.
\eqa
Note that this transformation does not respect the original `complex conjugacy' of $\psi_i^a$ and $\psi_i^{a*}$.
However it is legitimate 
because $\psi_1 \equiv \psi_i^a$ and $\psi_2 \equiv \psi_i^{a*}$ 
can be regarded as two independent Grassmann variables.
The partition function in the transformed fields becomes,
\bqa
Z & = & \int \sum_z \Pi_{i,a} \left(
d \psi_i^{a'*} 
d \psi_i^{a'} 
\right)
e^{-S},
\eqa
with the action 
\bqa
S & = &  - \sum_a \sum_i \psi_i^{a'} \psi_i^{a'*} \nn
&& - \frac{\kappa}{4} \sum_{a<b} \sum_{<i,j>} 
\left[  \psi^{a' *}_i \psi^{a' }_j \psi^{b'*}_j \psi^{b'}_i   e^{i \frac{\pi}{2} (z_i - z_j)}
+ c.c. \right] \nn
\label{f1s2}.
\eqa
In the following we will drop $'$ sign in $\psi^{a' }_i$.
The procedure of decomposing the $Z_2$ spin and
the fermionic variables 
is exactly the same as the bosonic case in the previous section.
The final low energy action is obtained to be two decoupled theories,
\bqa
S & = & 
 \sum_a \sum_i \psi_i^{a*} \psi_i^{a} 
-t \sum_a \sum_{<i,j>} \Bigl[ 
  e^{i a_{ij}} \psi^{a *}_i \psi^{a }_j + c.c.
\Bigr] \nn
&& - t_z \sum_{<i,j>} z_i z_j.
\label{f1_action}
\eqa

The Ising variable can have disordered and ordered phases as discussed in Sec. III. B.
In the following we focus on the fermions coupled with the compact U(1) gauge field.
Note that this fermion is spinless (structureless).
The spin-statistics theorem does not apply here because there is no full Lorentz invariance in lattice.
It is in the long wavelength limit much larger than the lattice scale
where the spin-statistics theorem requires half integer spin for the fermion
because of the full Lorentz invariance.
In the following we will see how the fermion acquires spin at low energy.
Similarly to the bosonic case,
we focus on one species of fermion by
integrating out the rest of fermions.
We obtain the effective action for the gauge field,
\bqa
S
& = &        
\sum_i \psi_i^{*} \psi_i 
 -t \sum_{<i,j>}
\left[ e^{i a_{ij}} \psi^{*}_i \psi_j + c.c.  \right] \nn             
&& + \sum_{C} \frac{1}{2g_C^2} \cos b_C.
\label{f1_u1}
\eqa
$b_C \equiv \oint_C a$ is the flux associated with the loop $C$. 
$g_C^2 \sim \frac{1}{(N-1) t^{L(C)}}$ is the generalized gauge coupling
for the non-local gauge action with $L(C)$, the length of the loop $C$.
Note that the sign of the gauge action is opposite to the one obtained from the bosonic action (\ref{f1_u1}).
The difference in sign comes from the difference in the statistics of the integrated matters.
This difference in statistics makes the difference in the low energy spin structure, 
confirming the spin-statistics theorem.

In the small $t$ limit, the unit plaquette term is most important 
in the gauge kinetic energy term.
Because of the opposite sign in the gauge kinetic energy term,
the saddle point of (\ref{f1_u1}) is `$\pi$-flux' phase where every space-time plaquette encloses flux $\pi$.
Now we derive the low energy theory for the fermion in $2+1D$.
Generalization to higher dimension is straightforward.
For the `$\pi$-flux', it is convenient to work in a unit cell 
which is doubled in both the $x$ and $y$ directions, but not 
the $z$ direction, with the gauge choice
\bqa
a_{r+x, r} & = & i, \nn
a_{r+y, r} & = & (-1)^{r_x + r_y}, \nn
a_{r+z, r} & = & i (-1)^{r_x + r_y}.
\eqa
The above `$\pi$-flux' has been used in lattice gauge theory to put Dirac fermion in lattice using only one-component fermion\cite{SUSSKIND}.
The difference from the Ref. \cite{SUSSKIND} is that in our case the $z$ direction is the imaginary time direction.
More importantly in our case the `$\pi$-flux' is the dynamical consequence while it is assumed in Ref. \cite{SUSSKIND}.
In the `$\pi$-flux' phase the action for the remaining fermion becomes
\bqa
S & = &   
\sum_I \sum_{A=1}^4 \psi_{IA}^* \psi_{IA}  \nn
&&
- t \sum_{I} \Bigl[ 
i   \psi^{*}_{I2} \psi_{I1}  
-   \psi^{*}_{I3} \psi_{I2}  
- i \psi^{*}_{I4} \psi_{I3}  
+   \psi^{*}_{I1} \psi_{I4}   \nn
&&
+ i \psi^{*}_{I+x 1} \psi_{I 2}  
+ i \psi^{*}_{I+x 4} \psi_{I 3}  
+   \psi^{*}_{I+y 2} \psi_{I 3}  
-   \psi^{*}_{I+y 1} \psi_{I 4}   \nn
&&
+ i \psi^{*}_{I+z 1} \psi_{I 1}  
- i \psi^{*}_{I+z 2} \psi_{I 2}  
+ i \psi^{*}_{I+z 3} \psi_{I 3}  
- i \psi^{*}_{I+z 4} \psi_{I 4}  
\Bigr] \nn
&& - c.c.,
\eqa
where $I$ is the unit cell index 
and $A = 1,2,3,4$ labels the sites within a unit cell.
Introducing four-component `spinor',
$\bar \Psi_{I}  =  (
\psi^*_{I 1}, 
\psi^*_{I 2}, 
\psi^*_{I 3}, 
\psi^*_{I 4} )$, 
$ \Psi_{I}^T  =  (
\psi_{I 1},
\psi_{I 2},
\psi_{I 3},
\psi_{I 4}  )$,
the action is written
\bqa
S & = &
 \sum_{I} 
\bar \Psi_{I} \left[
 \hat I 
- t
\left(
\begin{array}{cccc}
-2i \Delta_z & -i \Delta_x & 0 & \Delta_y \\
-i \Delta_x & 2i \Delta_z & -\Delta_y & 0 \\
0 & \Delta_y & -2i \Delta_z & -i \Delta_x \\
-\Delta_y & 0 & -i \Delta_x & 2i \Delta_z
\end{array}
\right)
\right]
\Psi_{I} \nn
&& + O(\Delta^2),
\eqa
where 
$\hat I$ is identity matrix and
$\Delta_p \Psi_I = \Psi_{I+p}-\Psi_{I}$.
We rescale the coordinates
$\Delta_x  \rightarrow  \partial_x$,
$\Delta_y  \rightarrow  \partial_y$, 
$2\Delta_z  \rightarrow  \partial_z$
in the continuum limit.
Introducing a new basis
\bq
\Psi  =  U \Psi^{'},
\bar \Psi  =  \bar \Psi^{'} U^\dagger
\eq
with
\bq
U = 
\frac{1}{\sqrt{2t}}
\left(
\begin{array}{cccc}
0 & -i  & 1 & 0 \\
i & 0  & 0 & 1 \\
0 & -i  & -1 & 0 \\
i & 0  & 0 & -1
\end{array}
\right),
\eq
and allowing the fluctuation of the gauge field,
we obtain the canonical Dirac Lagrangian,
\bqa
{\cal L} 
& = & 
\bar \Psi^{'} \left[ i \gamma_\mu ( \partial_\mu - i a_\mu ) + \frac{1}{t} \right] \Psi^{'}
+ \frac{1}{g^2} f_{\mu \nu}^2,
\eqa
where
\bqa
\gamma_\mu & = &
\left(
\begin{array}{cc}
-\sigma_\mu & 0   \\
0 & \sigma_\mu  
\end{array}
\right)
\eqa
with the Pauli matrices $\sigma_\mu$ for $\mu=1,2,3$.
This is the 4-component Dirac fermion with mass $\frac{1}{t}$.
Note that we obtained 4-component Dirac fermion even though
the minimal Dirac fermion has two component in (2+1)-dimension\cite{8COMP}.
This is because of the parity symmetry in our theory.
A mass term involving only the 2-component Dirac fermion would have broken parity symmetry\cite{APPEL}.

In (2+1) dimension only confinement phase can occur for the compact U(1) gauge theory with the massive Dirac fermions.
However the mass of the fermion decreases as $t$ increases.
Thus it is possible that the Dirac fermions become massless at a critical point, 
stabilizing the deconfinement phase\cite{HERMELE}.
If $t$ increases beyond the critical point, the off-diagonal excitons condense.
This corresponds to pair condensation of the Dirac fermions, 
which generates generalized mass term $\psi^{a*} M_{ab} \psi^b$.
To identify massless excitations in the condensed phase,
we decompose the phase of the off-diagonal exciton into two parts as
$\theta^{ab}_i = \theta^{(0)ab}_i + ( \phi^a_i - \phi^b_i )$.
Here $\theta^{(0)ab}_i$ parameterizes the fluctuations 
which involves the change in the sum of phases
$\theta^{a_1 a_2}_i + \theta^{a_2 a_3}_i + ...  + \theta^{a_{2n} a_1}_i$
and $\phi^a_i$, the fluctuations which does not change the sum.
Because of the frustration in Eq. (\ref{f1p}),
there are flat directions in both $\theta^{(0)ab}_i$ and $\phi^a_i$ classically.
Although not shown here, quantum fluctuations lift the degeneracy in $\theta^{(0)ab}_i$ 
because of singularities in the space of classical vacuum.
This leaves only the $(N-1)$ massless Goldstone modes.
The massless spectrum of the exciton condensed phase with $\tilde K_4 > 0$ 
is the same as that of the Higgs phase with $\tilde K_4 < 0$.
Thus we expect that the two phases are continuously connected. 
It is, of course, a possibility that the two phases are separated by a phase transition.

It is interesting to note that the massless Dirac fermions can arise at 
the quantum critical point of the bosonic matrix model.
The critical point is described by the algebraic spin liquid\cite{RANTNER}.
It corresponds to the deconfined quantum criticality\cite{SENTHIL04}
where massless modes at critical point stabilizes deconfinement phase.
This is the first example where fractionalized fermions occur at the critical point
as opposed to fractionalized bosons\cite{SENTHIL04}.

The frustration in the interaction is closely related to the fermionic nature of fractionalized particle.
And the spin follows the fermionic statistics.
In this sense both the emergent fermionic statistics and the spin are the results of the frustration in the bosonic matrix model.
In (3+1)-dimension the confinement to deconfinement phase transition can occur.
For small $t$, the confining phase arises where the fermions are confined by the U(1) gauge field.
For large $t$, the massless U(1) gauge field (photon) and gapped fermions emerges as low energy excitations.
The schematic phase diagram in (3+1)-dimension is shown in Fig. \ref{fig:phase} (b).

It must be emphasized that the fractionalized fermion is a different object from the original fermion of the microscopic Hamiltonian (\ref{micro}).
The original fermion is spinless and has no internal gauge charge, but it carries electromagnetic (EM) charge.
The fractionalized fermion has spin and charge for the internal gauge field. 
The only quantum number they have in common is the flavor index.
Moreover the fractionalized fermion is particle-hole symmetric, 
while the original fermions are in general not in that $m^c \neq m^d$.

\section{Duality mapping of the compact U(1) gauge theories with massive matter fields}

In the next section, we will describe the matrix models in dual representation. 
The goal of the dual description is to provide a more vivid picture of fractionalization 
and the emergent gauge field by describing the same phenomena in terms of the original exciton language.
Before doing that, in this section we will review the dual representation of 
the compact U(1) gauge theory coupled with matter fields.
This will help to identify what plays the role of the fractionalized particle and the gauge field in the original exciton picture which will be discussed in the next section.

\subsection{bosonic matter}

In this section we discuss the dual representation of the compact U(1) gauge theory coupled
with bosonic matter field.
The action (\ref{b1_u1}) is rewritten 
\bqa
S & = & -\frac{t}{2} \sum_{i} \sum_{\mu} \left( \cos ( \phi(i+\mu) - \phi(i) - a_{\mu}(i) ) -1 \right) \nn
 && - \sum_C \frac{1}{2g_C^2} \left( \cos  b_C -1 \right).
\label{gb_s}
\eqa
Here $i$ is lattice site and $\mu$ refers to $D$ positive direction of link. 
$t$ is the phase stiffness of boson 
and $g_C$, the generalized gauge coupling associated with the flux through a 
loop $C$.
Each loop is assigned a unique orientation and $b_C = \oint_C a$.
The choice of orientation does not matter because of the $\cos  (b_C)$.
The summation of $C$ in the second term in Eq. (\ref{gb_s}) is over 
all possible closed loops.
If only the smallest plaquette is allowed for the gauge kinetic term,
Eq. (\ref{gb_s}) reduces to the well known Abelian-Higgs model\cite{KOGUT}.
Using the Villain approximation, we write the partition function,
\bqa
Z & = & \int Da \int D \phi 
\sum_{p_{\mu}(i) = -\infty}^\infty  \sum_{q_C = -\infty }^\infty   \nn
&& \exp \Bigl( 
-\frac{t}{4} \sum_{i,\mu} \{ \phi(i+\mu) - \phi(i) - a_{\mu}(i) - 2\pi p_{\mu}(i) \}^2   \nn
&& - \sum_C \frac{1}{4g_C^2} \{ b_C - 2 \pi q_C \}^2
\Bigr).
\eqa
Hubbard-Stratonovich transformation for the quadratic terms leads to
\bqa
Z 
& = & 
\int Da \int D \phi 
\int D l_{\mu} \int D n_C
\sum_{p_{\mu}(i), q_C }   \nn
&& \exp \Bigl( 
-\frac{1}{t} \sum_{i,\mu} (l_{\mu}(i))^2    \nn
&& -i \sum_{i,\mu} l_{\mu}(i) \{ \phi(i+\mu) - \phi(i) - a_{\mu}(i) - 2\pi p_{\mu}(i) \}   \nn
&& -\sum_C g_C^2 n_C^2
-i \sum_C n_C \{ b_C - 2 \pi q_C \} 
\Bigr).
\eqa
From the identity
$\sum_{n = -\infty}^{\infty} e^{2\pi i n x} = \sum_{n = -\infty}^{\infty} \delta(x - n)$
the partition function becomes the sum over integer fields,
\bqa
Z
& = & 
\int Da \int D \phi 
\sum_{l_{\mu}(i), n_C }   
 \exp \Bigl( 
-\frac{1}{t} \sum_{i,\mu} (l_{\mu}(i))^2    \nn
&& -i \sum_{i,\mu} l_{\mu}(i) \{ \phi(i+\mu) - \phi(i) - a_{\mu}(i) \}   \nn
&& - \sum_C g_C^2 n_C^2
-i \sum_C n_C b_C 
\Bigr).
\eqa
Integrating out $\phi$ and $a$ we obtain 
the constraint for the current conservation of boson
and the flux conservation of the electric flux line,
\bqa
&& Z  = 
\sum_{l_{\mu}(i), n_C }   
 \exp \Bigl( 
-\frac{1}{t} \sum_{i,\mu} (l_{\mu}(i))^2   
- \sum_C g_C^2  n_C^2 \Bigr) \times \nn
&& \left( \Pi_{i} \delta_{\sum_\mu \Delta_\mu l_{\mu}(i), 0}  \right)
\left( \Pi_{i,\mu} \delta_{  l_\mu(i) - \sum_{C^{'} \ni (i,\mu)} 
S(C^{'},i,\mu) n_{C^{'}}, 0} \right). \nn
\label{gbz}
\eqa
Here $l_{\mu}(i)$ is integer defined on link.
Nonzero value of it at a given line $(i,\mu)$ represents that 
the link is a part of world lines of boson(s). 
Negative value represents world line of anti-boson.
The first constraint in the second line of (\ref{gbz}) implies that
the world line of boson should form closed loop.
This is the result of current conservation of boson.
$n_C$ is an integer valued `field' assigned to 
the loop $C$, i.e., it is defined on the space of loops,
the set of all possible closed loops.
The typical size of the loop is given by the inverse mass of the integrated matter field.
It is convenient to think of the loop as a boundary of a surface.
Then nonzero value of $n_C$ implies that
the surface enclosed by $C$ is a part of a world sheet passing through the surface.
The value with $n_C > 1$ is regarded as multiple world sheets and 
negative value, a world sheet of opposite orientation.
The sign $S(C,i,\mu)$ which appears in the second constraint in Eq. (\ref{gbz})
is $1$ ($-1$) if the orientation of $C$ is along (against) the vector $\mu$ associated
with the $(i,\mu)$.
This constraint implies that such world sheet cannot end by itself.
In the absence of boson current ($l_{\mu}(i) = 0$)
the end of one world sheet should be joined by the end of another world sheet
of same orientation.
This means that they should form oriented closed surface in space-time.
In the presence of boson current ($l_{\mu}(i) \neq 0$) the world sheet 
can end on the world line of boson.
In lattice gauge theory, $n_C \neq 0$ represents 
the presence of quantized electric flux on that link.
Thus the bosons serve as sources of the electric flux lines.
The second constraint in Eq. (\ref{gbz}) is the flux conservation law.
Note that the world sheet is not well-defined in the lattice scale 
because of the uncertainty in determining a unique surface enclosed by a loop.
We can define a smoothly varying surface only in the long distance scale 
larger than the typical length of the loops.
This is attributed to the fact that gauge theory emerges
only in the energy scale lower than the mass of the integrated
matter field.

\subsection{fermionic matter}

Here we consider the compact U(1) gauge theory coupled with fermion in (\ref{f1_u1}),
\bqa
S & = & 
-t \sum_{i} \sum_{\pm \mu} 
e^{i a_\mu(i)}
\psi_{i+\mu}^*  \psi_{i}        \nn
&& + \sum_{i} \psi_{i}^*  \psi_{i}       
- \sum_C \frac{1}{2g_C^2}  \left( - \cos  b_C  - 1 \right).
\label{gfs}
\eqa
Here $\pm \mu$ denote the nearest neighbor link in all $2D$ directions and
$a_{-\mu}(i+\mu) = -a_\mu(i)$.
Note that the sign of the gauge kinetic energy term is opposite to that of (\ref{gb_s}).
We introduce an additional phase $\pi$ in the flux term 
to apply the usual Villain approximation and the Hubbard-Stratonovich transformation,
\bqa
e^{\frac{1}{2g_C^2} ( -\cos b_C - 1 )} & = &
\sum_{n_C }   
e^{
 -\sum_C  g_C^2 n_C^2 
-i \sum_C n_{C}  ( b_C- \pi)  }. \nn
\label{gaugeg}
\eqa
For the matter field we expand in $t$,
\bqa
Z_\psi & \equiv &
\int 
D\psi^* 
D\psi 
e^{ 
t \sum_{i} \sum_{\pm \mu} e^{i a_\mu(i)}
\psi_{i+\mu}^*  \psi_{i}       
- \sum_{i} \psi_{i}^*  \psi_{i}       
}  \nn
& = & 
\sum_{n=0}^\infty
\frac{1}{n!}
\left<  
\left(
t \sum_{i} \sum_{\pm \mu} e^{i a_\mu(i)}
\psi_{i+\mu}^*  \psi_{i}       
\right)^n
\right>,
\eqa
where $< A > \equiv \int D\psi^* D\psi A e^{- \sum_{i} \psi_{i}^*  \psi_{i}}$.
Because of the conservation of matter current,
only loop of fermion world lines contribute to the partition function,
\bqa
Z_\psi & = &
\sum_{m=0}^\infty
(-1)^m
\sum_{C_1, ..., C_m}
t^{\sum_{l=1}^m L(C_m)}
e^{i \sum_{l=1}^m \oint_{C_l} a
}. \nn
\label{gaugem}
\eqa
Here $m$ is the number of fermion loops where each loop is represented as $C_m$.
There is a distinction of orientation for each fermion loop which
is determined by the direction of fermion current.
Unlike the bosonic case there is no ambiguity in counting the number of loops.
This is because two loops cannot intersect owing to the anti-commutativity $\psi_i^2 = 0$.
$L(C_m)$ is the length of each loop
and $\oint_{C_l} a$ is the U(1) gauge flux penetrating the loop.
The alternating sign with the number of fermion loops originates from 
the anticommutativity of the Grassmann field.
Using (\ref{gaugeg}) and (\ref{gaugem}) and integrating out
the gauge field we obtain
\bqa
&& Z
 = 
\sum_{l_\mu(i) = 0, \pm 1 }   
\sum_{n_C }   
(-1)^{m(l_\mu)}
e^{ -\sum_C ( g_C^2 n_C^2 - i \pi n_C)  }
t^{\sum_{(i,\mu)} |l_{\mu}(i)|} \nn
&& \times
\left( \Pi_{i} \delta_{\sum_\mu \Delta_\mu l_{\mu}(i), 0}  \right)
\left( \Pi_{i,\mu} \delta_{ l_\mu(i) -  \sum_{C^{'} \ni (i,\mu)} S(C^{'},i,\mu) n_{C^{'}} , 0} \right). \nn
\label{gfz}
\eqa
Here $m(l_\mu)$ is the number of fermion loops for a given configuration of $l_\mu(i)$.
This is similar to the action with the bosonic matter in (\ref{gbz}) except that
$l_{\mu}(i)$ can have only $0$ and $\pm 1$,
$e^{-1/t}$ is replaced with $t$
and there is a factor $(-1)$ for each fermion loop and $n_C$.
The minus sign for each $n_C$ originates from the opposite sign in the
gauge kinetic energy term.

\section{Dual approach to the Emergent U(1) gauge theory in the matrix models}

In this section we dualize the three different theories of (\ref{category}) in the world line representation of exciton.
The fractionalization, the emergence of photon in the Coulomb phase and the statistics of the fractionalized particle will be discussed in the exciton language.

\subsection{Duality description of fractionalized boson and U(1) gauge field}

\begin{figure}
        \includegraphics[height=7cm,width=9cm]{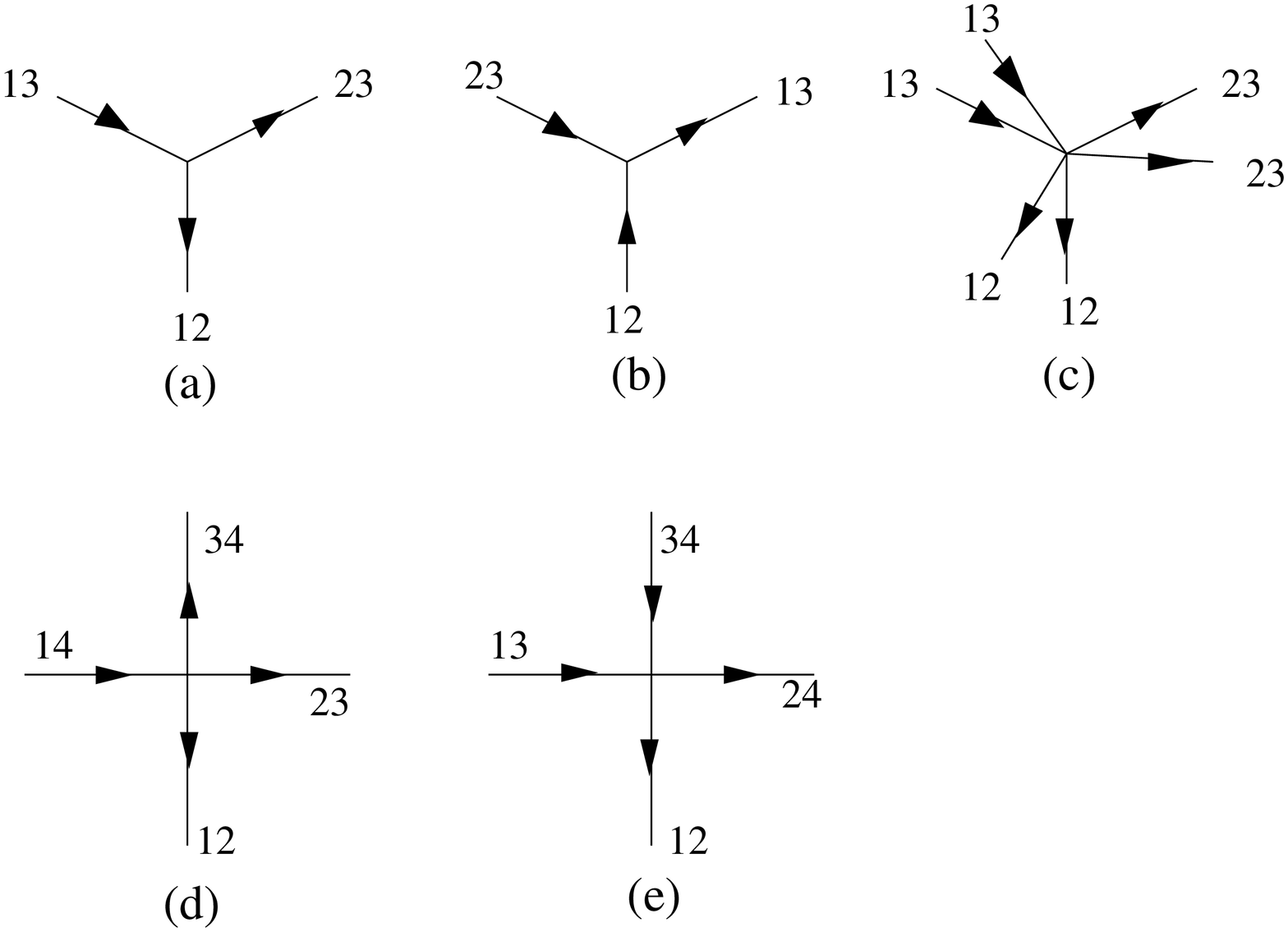}
\caption{
Examples of vertices in the world line representation of exciton
with
(a) $V^{123}=1$,
(b) $V^{213}=1$,
(c) $V^{123}=2$,
(d) $V^{1234}=1$ and
(e) $V^{1243}=1$.
An incoming $ab$-exciton is same as an outgoing $ba$-exciton because $\chi$ is Hermitian.
The direction of the exciton world line attached to vertices are drawn
with the reference of $ab$-exciton with $a<b$.
}
\label{fig:vertices}
\end{figure}

\begin{figure}
        \includegraphics[height=7cm,width=9cm]{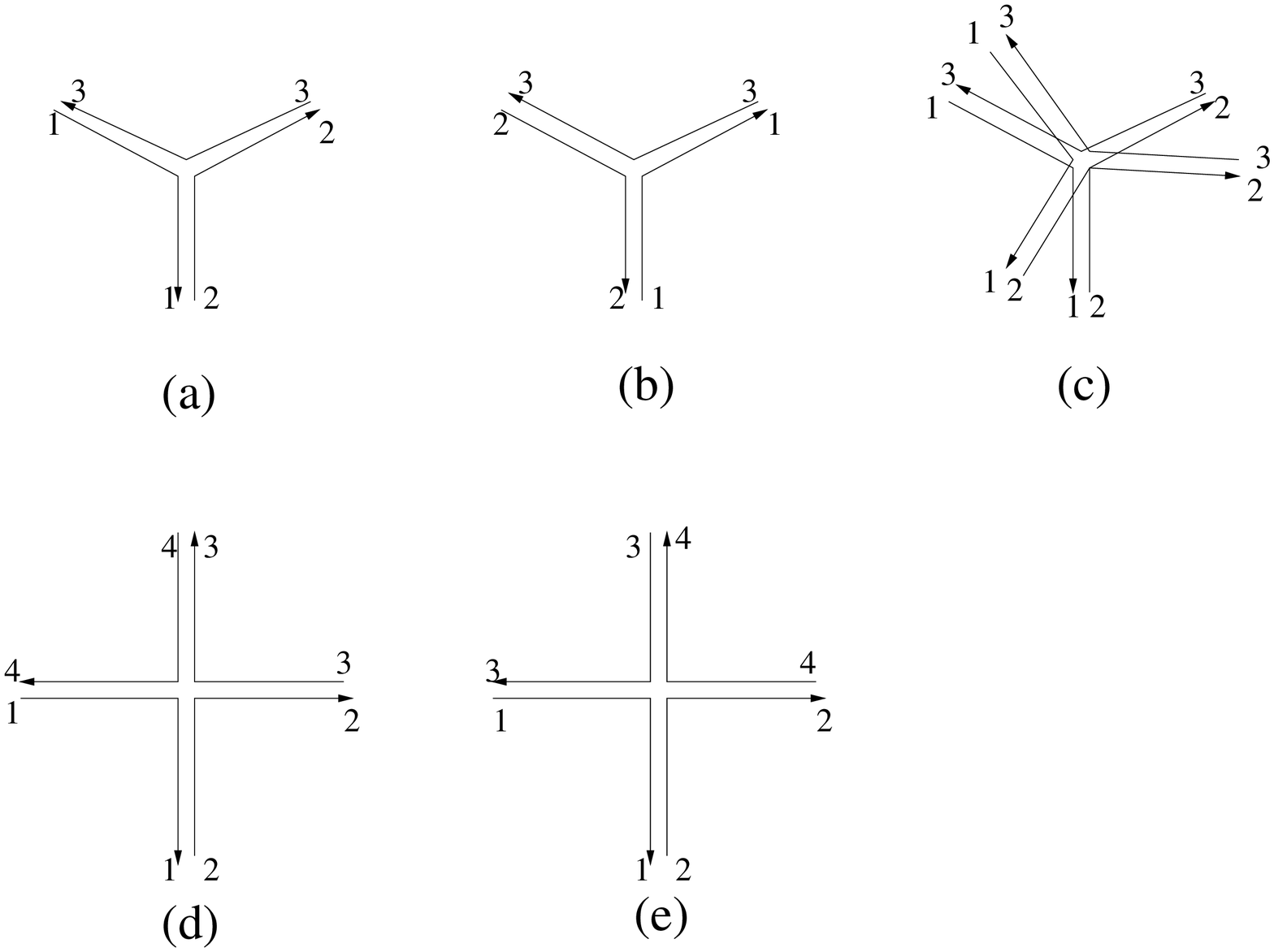}
\caption{
Double line representation of the same vertices shown in Fig. \ref{fig:vertices} with
(a) $V^{123}=1$,
(b) $V^{213}=1$,
(c) $V^{123}=2$,
(d) $V^{1234}=1$ and
(e) $V^{1243}=1$.
The world line of exciton with flavor $ab$ is denoted as two lines where
one line of flavor $a$ moves in the same direction as the
exciton world line and the other line of flavor $b$ in the opposite direction.
}
\label{fig:vertices_double}
\end{figure}

Consider the effective action for off-diagonal phases of the Hermitian matrix with $\tilde K_n < 0$,
\begin{widetext}
\bqa
S  & = &
-\frac{\kappa}{2} \sum_{i} \sum_{\mu} \sum_{a<b} 
\left[ \cos \left( \theta^{ab}(i+\mu) - \theta^{ab}(i) \right)  -1 \right]  \nn
&& - \sum_{n=3}^{\infty} \frac{\kappa_n}{2} \sum_{i} \sum_{[a_1,a_2,...,a_n]}
\left[
f^{[a_1,a_2,...,a_n]}
\cos \left( \sum_{l=1}^{n} \theta^{a_l a_{l+1}}(i) \right) - 1 \right].
\eqa
Here $\theta^{ab} = - \theta^{ba}$  is the phase of the Hermitian matrix.
We defined $a_{n+1} \equiv a_1$ for notational simplicity.
$\kappa = 2 K \chi_o^2$
and 
$\kappa_n = 4 n | \tilde K_n | \chi_o^n$.
The summation $[a_1,a_2,...,a_n]$ is over all $a_1$, ..., $a_n$ modulo cyclic permutation 
and the reverse of the order of indices.
No repeated consecutive indices are allowed, that is, $a_i \neq a_{i+1}$.
$f^{[a_1,a_2,...,a_n]}$ is a symmetry factor which compensate the double counting in the
cyclic permutation and the reverse of the indices.
For example, 
$f^{[1,2,3]} = f^{[1,2,3,4]}  = 1$ and
$f^{[1,2,3,1,2,3]} = \frac{1}{2}$, etc.
With the Villain approximation for both the kinetic energy term and the
interaction term, the partition function becomes
\bqa
Z & = & \int D \theta
\sum_{p_{\mu}^{ab}(i) = -\infty}^\infty
\Pi_{n=3}^\infty
\left( \sum_{q_{[a_1,..,a_n]}(i) = -\infty }^\infty \right)  
 \exp \Bigl[
-\frac{\kappa}{4} \sum_{i,\mu,a<b} \{ \theta^{ab}(i+\mu) - \theta^{ab}(i) - 2\pi p_{\mu}^{ab}(i) \}^2   \nn
&& - \sum_{n=3}^{\infty} \frac{\kappa_n}{4} \sum_{i} \sum_{[a_1,a_2,...,a_n]}
f^{[a_1,a_2,...,a_n]}
\{ \sum_{l=1}^{n} \theta^{a_l a_{l+1}}(i)  - 2\pi q_{[a_1,..,a_n]}(i)  \}^2
\Bigr].
\eqa
Introducing two Hubbard-Stratonovich fields $l^{ab}_{\mu}(i)$ and $V^{[a_1 ... a_n]}(i)$
and performing the summation over $p_{\mu}^{ab}(i)$ and $q_{[a_1,..,a_n]}(i)$,
the partition function becomes sum over integer numbers
for the exciton current and interaction vertex,
\bqa
Z
& = &
\sum_{l_{\mu}^{ab}(i), V^{[a_1...a_n]}(i) }   
 \exp \Bigl( 
-\frac{1}{\kappa} \sum_{i,\mu,a<b} (l_{\mu}^{ab}(i))^2   
- \sum_{n=3}^{\infty} \sum_{i} \sum_{[a_1,a_2,...,a_n]} \frac{1}{\kappa^{[a_1...a_n]}}  (V^{[a_1 ... a_n]}(i))^2
\Bigr) \times \nn
&& \int D \theta  
\exp \Bigl(
-i \sum_{i,\mu,a<b} l_{\mu}^{ab}(i) \{ \theta^{ab}(i+\mu) - \theta^{ab}(i) \}   
-i \sum_{n=3}^{\infty} \sum_{i} \sum_{[a_1,a_2,...,a_n]} 
  V^{[a_1 ... a_n]}(i)  \sum_{l=1}^{n} \theta^{a_l a_{l+1}}(i) 
\Bigr), \nn
\label{db1z}
\eqa
\end{widetext}
where $\kappa^{[a_1...a_{2n}]} = \kappa_{2n} f^{[a_1,a_2,...,a_{2n}]}$.
Here $l^{ab}_\mu(i)$ represents current of $ab$ excitons and
$V^{[a_1 ... a_n]}(i)$, vertex of interaction.
The free energy cost of unit current in a link is $\frac{1}{\kappa}$
and that of unit vertex at a site is $\frac{1}{\kappa^{[a_1...a_n]}}$.
Integration over the exciton phase leads to current conservation condition,
\bq
\sum_\mu \Delta_\mu l^{ab}_{\mu}(i) - \sum_{n} \sum_{a_3,..,a_n} (V^{[ab a_3...a_{n}]} - V^{[ba a_3...a_{n}]} )= 0.
\label{const}
\eq
The constraint  (\ref{const})
implies that the vertex $V^{[a_1 a_2 . ..a_{n}]}$ is a source or sink of the currents.
Exciton current of each flavor is conserved if $V^{[a_1 a_2 . ..a_{n}]}=0$ for all $[a_1 a_2 . ..a_{n}]$. 
If $V^{[a_1 a_2 . ..a_{n}]}(i) \neq 0$, the site $i$ becomes a source for currents
$j^{a_1 a_2}$,..., $j^{a_i a_{i+1}}$, ..., $j^{a_n a_{1}}$
and a sink for the currents
$j^{a_2 a_1}$,..., $j^{a_{i+1} a_{i}}$, ..., $j^{a_1 a_{n}}$.
Examples of the vertices are shown in Fig. \ref{fig:vertices}(a)-(e). 
The vertices describe interaction among excitons.
One species of exciton is annihilated and other exciton is created at the vertex
as constituent fermion exchange their partners.
Even though the exciton current of each species is not conserved at vertex,
the flavor current is always conserved.
This becomes more transparent if we represent exciton world line in double lines (see Fig. \ref{fig:vertices_double}).

In the double line representation
the link with $l^{ab}_\mu(i) = 1$ can be viewed as two world lines of a particle of flavor $a$ and an anti-particle of flavor $b$.
Each single line will be interpreted as the world line of the emergent slave boson.
Since each flavor current is always conserved, the single line should form loop.
The single line should represent the world line of boson not fermion 
because the phase factor associated with the loop of single line is $1$.
This is consistent with the earlier discussion in Sec. III. A.
At this stage the boson is just a `book-keeping' tool for the exciton flavor. 
Whether this boson occurs as a low energy excitation or not is a dynamical question.
In the weak coupling limit ($\kappa_n << 1$) the energy cost of a vertex is very high.
Each exciton maintains its identity and the boson is confined within a single exciton for a long time $1/\kappa_n$.
The low energy physics is described by weakly interacting excitons.
This is the disordered phase of the off diagonal exciton condensate.
It is in the strong coupling limit ($\kappa_n >> 1$) where the bosons have chance to be liberated.
In the strong coupling limit the vertex is cheap and exciton completely looses its identity within the discretized time (distance) scale.
The scattering occurs so often that the boson is no longer confined within a single exciton.
The exciton is not a quasiparticle.
On the other hand the slave boson can propagate for much longer time (distance) than exciton as they keep changing their partner.
Then the slave boson emerges as low energy excitation of the system.
Some remarks are in order for the fractionalization. 
First, it does not mean that an exciton is really separated into two slave bosons.
Instead a boson is always part of some exciton. 
The point is that a boson is liberated from a particular exciton by frequently changing its partner in the strong coupling limit.
Then the boson (anti-boson) appears to be free in the medium of other anti-bosons (bosons).
(However the boson is not necessarily a quasiparticle because of the coupling to an emergent photon as will be discussed below.)
It is emphasized again that the drastic change from the exciton quasiparticle picture
to the fractionalization is due to the strong interaction.
Second, the concept of fractionalization can be useful in an intermediate energy scale even in the case 
where there is no fractionalization in the low energy limit. 
In the following we explain these in more detail.

\begin{figure}
        \includegraphics[height=12cm,width=8cm]{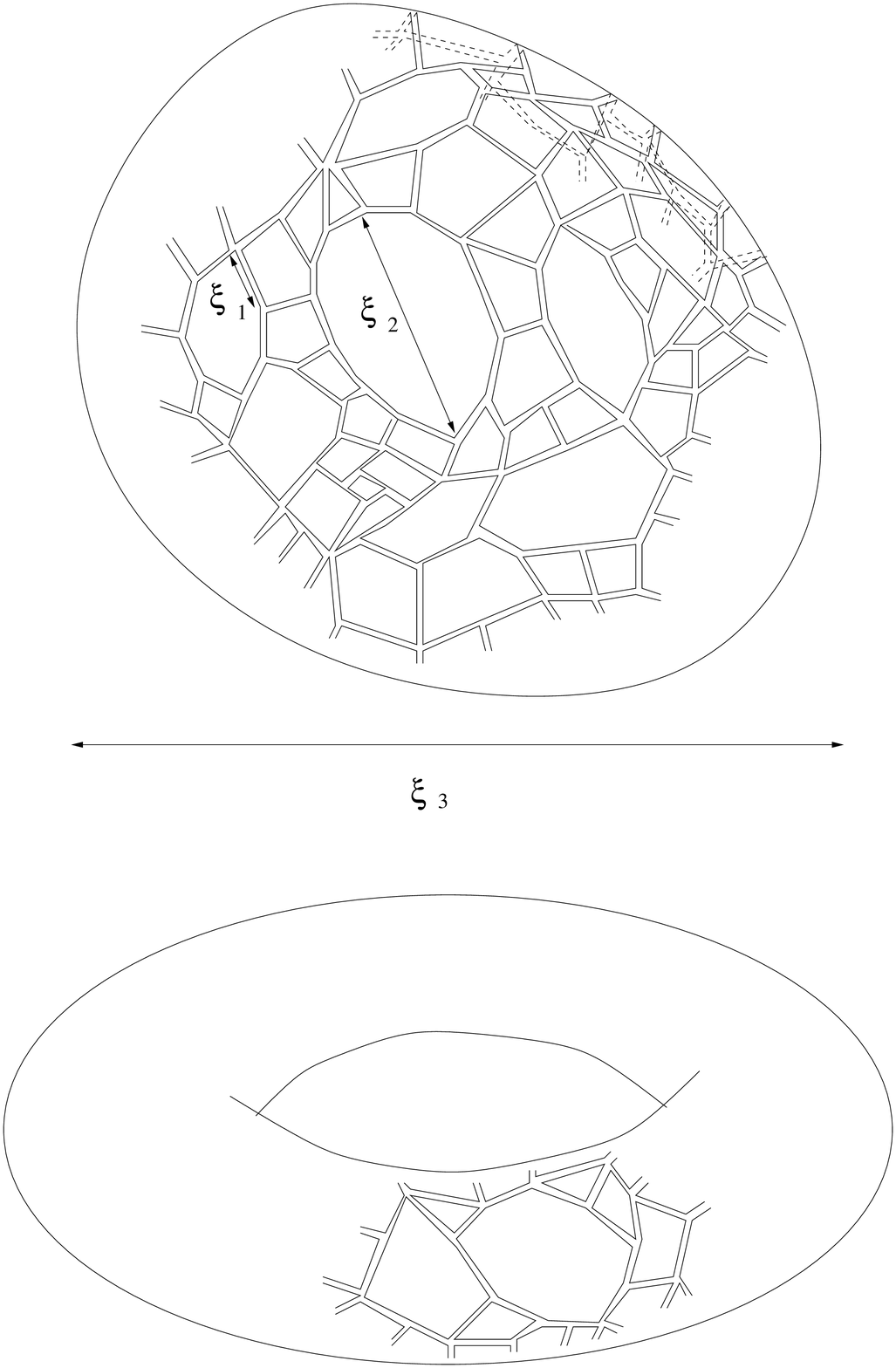}
\caption{
Sphere and torus formed by web of exciton world lines.
}
\label{fig:web}
\end{figure}

\begin{figure}
        \includegraphics[height=5cm,width=8cm]{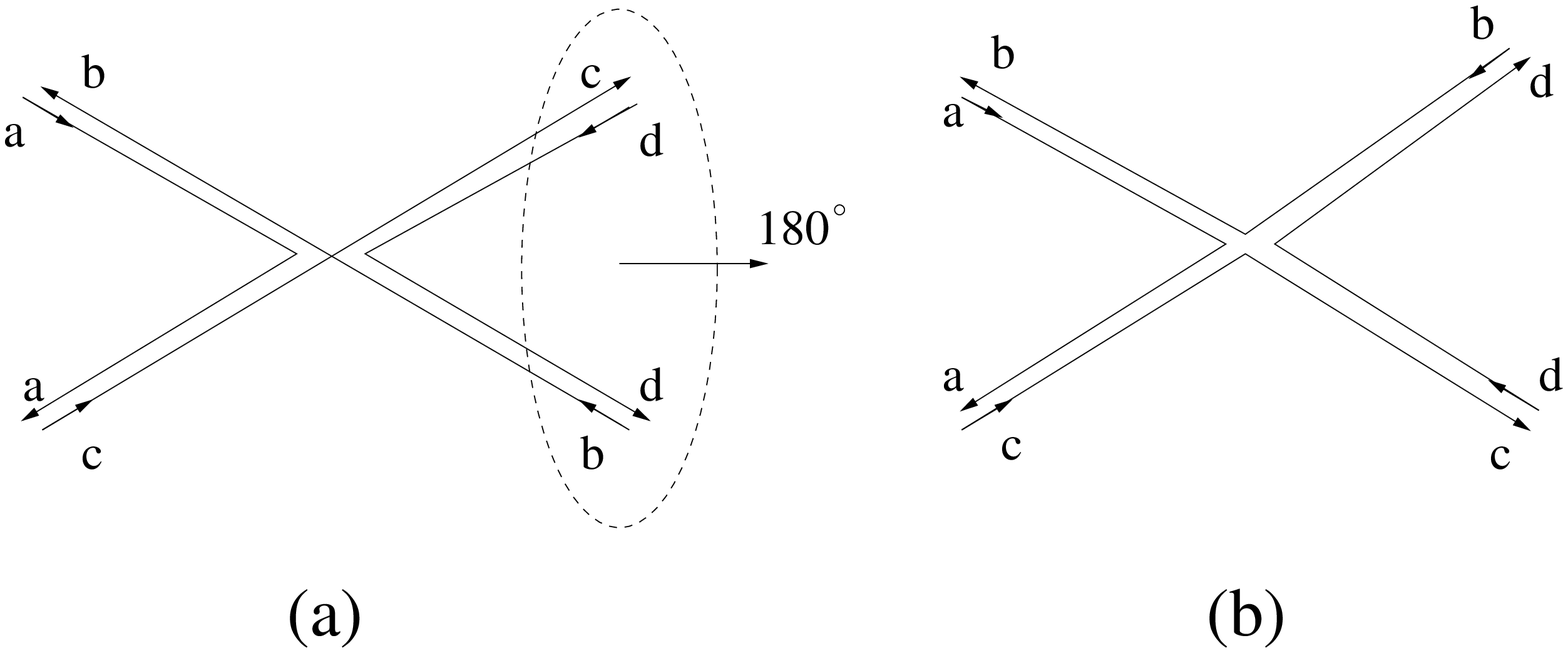}
\caption{
Crossing vertex as a twisted uncrossed vertex.
}
\label{fig:twist}
\end{figure}

First we consider what configurations contribute to partition function (\ref{db1z}). 
The world line of the slave boson forms closed loop and is accompanied with another line with opposite current.
The simplest configuration is two overlapped loops with current of flavor $a$ in one direction and $b$ in the other direction.
It describes a process where $ab$-exciton and $ba$-exciton pop out of vacuum and annihilate.
In this sense the $ab$ and $ba$-excitons are anti-particles to each other 
even though they are two independent excitons.
This is a consequence of the condensation of diagonal excitons. 
Two diagonal excitons ($aa$ and $bb$) in condensate collide to become two off-diagonal excitons ($ab$ and $ba$) vice versa.
The role of the phase coherent diagonal excitons is important even though it is hidden behind the dynamical stage.
Note that we have $\theta^{ab} = - \theta^{ba}$ because of the mass term in (\ref{mass}) which is generated by the phase coherent diagonal excitons.
General configurations consist of a collection of connected webs made of boson world lines (world line web).
Each world line web defines closed surfaces as depicted in Fig. \ref{fig:web}.
Many such closed surfaces may co-exist and interpenetrate each other.
It is noted that the closed surfaces made of the world line web are oriented.
This is because the boson current has direction and only two currents with opposite direction can share a link.
However, can we still define the orientation even if there is a crossing of flavor in a vertex ?
The answer is yes.
One crossing diagram is shown in Fig. \ref{fig:twist} (a).
The crossing vertex in Fig. \ref{fig:twist} (a) can be obtained from 
the non-crossing vertex in Fig. \ref{fig:twist} (b) 
through a continuous deformation involving rotation as is shown in the figure.
The oriented nature of the surface has important consequences for
the gauge group of the emergent gauge field
and the statistics of the slave particle 
as will be shown below.
Note that there are three independent length scales in the world line web.
The shortest scale is the length of double line segment ($\xi_1$) which corresponds to the life time of a single exciton.
The other is the size of the loop of single line ($\xi_2$) which is the life time of the slave boson.
Note that there is distribution in the size of the loops of single lines from the small loop
of size $\xi_1$ to the large loop whose size is much larger than $\xi_1$.
The smaller loops of size $\xi_1$ are filling the area between the large loops of size $\xi_2$.
Even though there exist large loops of single lines, the life time of a single exciton is kept to be $\xi_1$.
This is how the slave boson can emerge as low energy excitations. 
The largest length scale is the size of the bubble made of the web ($\xi_3$).
In general we have $\xi_1 < \xi_2 < \xi_3$.
In the following the confinement, Coulomb and Higgs phases are described in terms of the dynamics of the web. 

In the small $\kappa$ limit, all length scales are of order of lattice scale.
As $\kappa$ increases $\xi_2$ and $\xi_3$ increases. 
However $\xi_1$ remains almost same for fixed coupling constants ($\tilde \kappa_n >> 1$). 
For $\xi_2 << \xi_3$ the closed surface of the world line web is well-defined in the length scale $x \sim \xi_3$.
Thus the low energy theory reduces to the statistical problem of fluctuating surface.
The surface can be regarded as the world sheet of oriented closed string.
The string tension is determined by the density of exciton and $\kappa$.
In the gauge theory picture the string corresponds to electric flux line and the string tension, the gauge coupling.
The gauge group should be U(1).
This comes from the orientedness of the surface.
In contrast, in the $Z_n$ gauge theory there are lines where $n$ currents of same direction can meet. 
Then $n$ surfaces can join at these lines.
This leads to an unoriented surface. 
If $\xi_3$ is finite there is finite tension for the world sheet of electric flux line.
$\xi_3$ corresponds the confining scale.
How can we probe the confining gauge field ?
This can be done by examining the potential energy between the two slave bosons.

\begin{figure}
        \includegraphics[height=8cm,width=7cm]{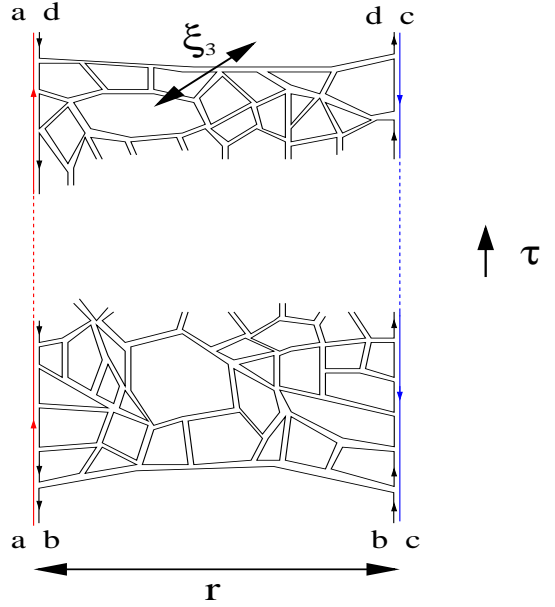}
\caption{
Propagation of two excitons separated by distance $r$ in the confinement phase.
The end points of the web are boson and anti-boson, and the web is the electric flux line connecting them.
$\xi_3$ is the length scale associated with the transverse fluctuations of the web.
}
\label{fig:conf}
\end{figure}

Consider creating two excitons of flavors $ab$ and $bc$ separated by distance $r$.
Then ask the probability amplitude for the two excitons to end up with 
excitons of flavors $ad$ and $dc$ after $\tau = \tau_1$.
We follow only the flavor of $a$ and $c$ by tracing out flavor $b$ and $d$ and measure
\begin{widetext}
\bqa
V(r) & \equiv & - \lim_{\tau_1 \rightarrow \infty} \frac{1}{\tau_1} 
\ln \sum_{b,d} \left<  
e^{-i( \theta^{ad}({\bf i},\tau_1) + \theta^{dc}({\bf i}+ {\bf r}, \tau_1) )}
e^{i( \theta^{ab}({\bf i},0) + \theta^{bc}({\bf i}+ {\bf r}, 0) )}
\right>,
\eqa
\end{widetext}
where ${\bf i}$ and ${\bf r}$ is the (D-1)-dimensional space vector 
and $\tau$, the imaginary time.
The world line of the $b$ and $d$ bosons will be present 
only for a finite time scale near $\tau = 0$ and $\tau_1$ respectively.
After an initial transient time, we are left with boson $a$ and anti-boson $c$ connected by the world line web as is shown in Fig. \ref{fig:conf}.
The end point of the web is the slave boson and the web is the electric flux line connecting them.
Thus we effectively measure the potential energy between the slave bosons.
If $\xi_2 << r << \xi_3$
two test bosons separated by $r$ are subject to Coulomb interaction,
$V(r) \sim \ln r$ for $D=3$ and 
$V(r) \sim -1/r^{D-3}$ for $D>3$. 
This is because  the world line web fluctuates wildly in this length scale. 
The linear potential, $V(r) \sim r$, sets in for $\xi_3 << r << \frac{\xi_3^2}{\xi_2}$  where the electric flux line is almost straight. 
For $r >> \frac{\xi_3^2}{\xi_2}$ the energy associated with the extended electric flux line exceeds the mass gap of slave boson. 
The boson and anti-boson pop out to screen the test bosons and create two separate excitons.
Then the potential becomes flat, $V(r) \sim constant$.
This is what we observe in the confinement phase of the U(1) gauge theory with massive bosons.
In the intermediate energy scale $1/\xi_3 << E << 1/\xi_2$ the effective theory is the emergent pure U(1) gauge field with almost gapless photon.
It is known that pure compact U(1) gauge theory is always confined in 2+1 dimension.
This means that $\xi_3$ is always finite and
in the lowest energy scale $E << 1/\xi_3$ every excitation is gapped.
The Coulomb phase is not a stable phase but can exist only as a cross-over.
In space dimension greater than $2$ the real Coulomb phase can arise even in the low energy limit.
In the following we discuss the Coulomb phase.

As $\kappa$ increases further, the bubbles grow and overlap.
The surfaces of two bubble can re-connect to form a large and a small bubbles.
In this way an infinitely large surface can form and the length scale $\xi_3$ diverges 
while $\xi_2$ remains finite.
The world line web condenses (but not the individual boson). 
The low energy excitation is the collective excitation of the condensed world line web.
In a given time slice, the intersection of the web surface with the time slice
forms strings.
In the deconfined phase, these strings are infinite in extent.
This is a space-time picture for the string-net condensation\cite{WEN}.
In the gauge theory language the electric flux lines are the strings which are
condensed, and the low energy collective excitation becomes the gapless `photon'.
This is the Coulomb (deconfinement) phase.

\begin{figure}
        \includegraphics[height=12cm,width=7cm]{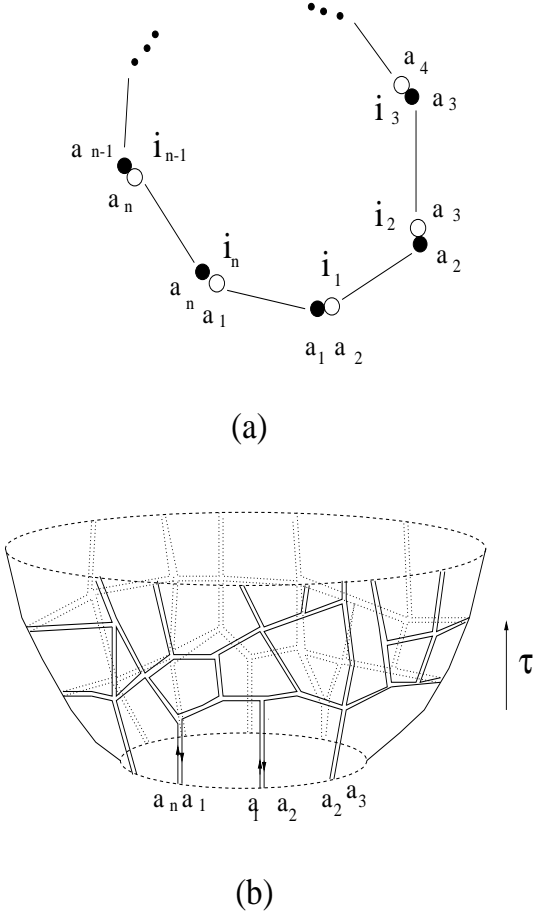}
\caption{
(a) Series of excitons created by the Wilson operator.
The solid (open) circle denotes flavor (anti-flavor) of exciton.
(b) World sheet spanned by the time evolution of the Wilson operator.
}
\label{fig:wilson}
\end{figure}

What is the massless `photon' in terms of original exciton ?
To answer the question we first construct the Wilson operator. 
The Wilson operator creates a closed electric flux line in the gauge theory picture.
Since the world sheet of electric flux line corresponds to world line web in the exciton picture,
the Wilson operator corresponds to creation operator of series of excitons.
However in order to create low energy excitations,
the operator needs to create excitons in a special order along the loop,
as we see from the following argument.
The world line web corresponding to low energy excitations need to have large number of loops. 
This is because each loop contributes $-ln N$ ($N$, the number of flavor) to the action.
From this consideration we identify the Wilson operator,
\bq
W_{i_1 i_2 ... i_n } \sim \sum_{a_1 a_2 ... a_n} 
e^{i ( 
\theta^{a_1 a_2}_{i_1} 
+ \theta^{a_2 a_3}_{i_2} 
+ ...
+ \theta^{a_n a_1}_{i_n}  )}.
\eq
This operator creates a series of exciton along a contour $C$ defined by the sites $i_1,...,i_n$.
The neighboring excitons have same flavor and anti-flavor (see Fig. \ref{fig:wilson} (a)).
As the series of excitons propagate in time they can readily form a web as is shown in Fig. \ref{fig:wilson} (b).
The correlation between flavors enables excitons to efficiently form a web maximizing the number of loops.

\begin{figure}
        \includegraphics[height=12cm,width=6cm]{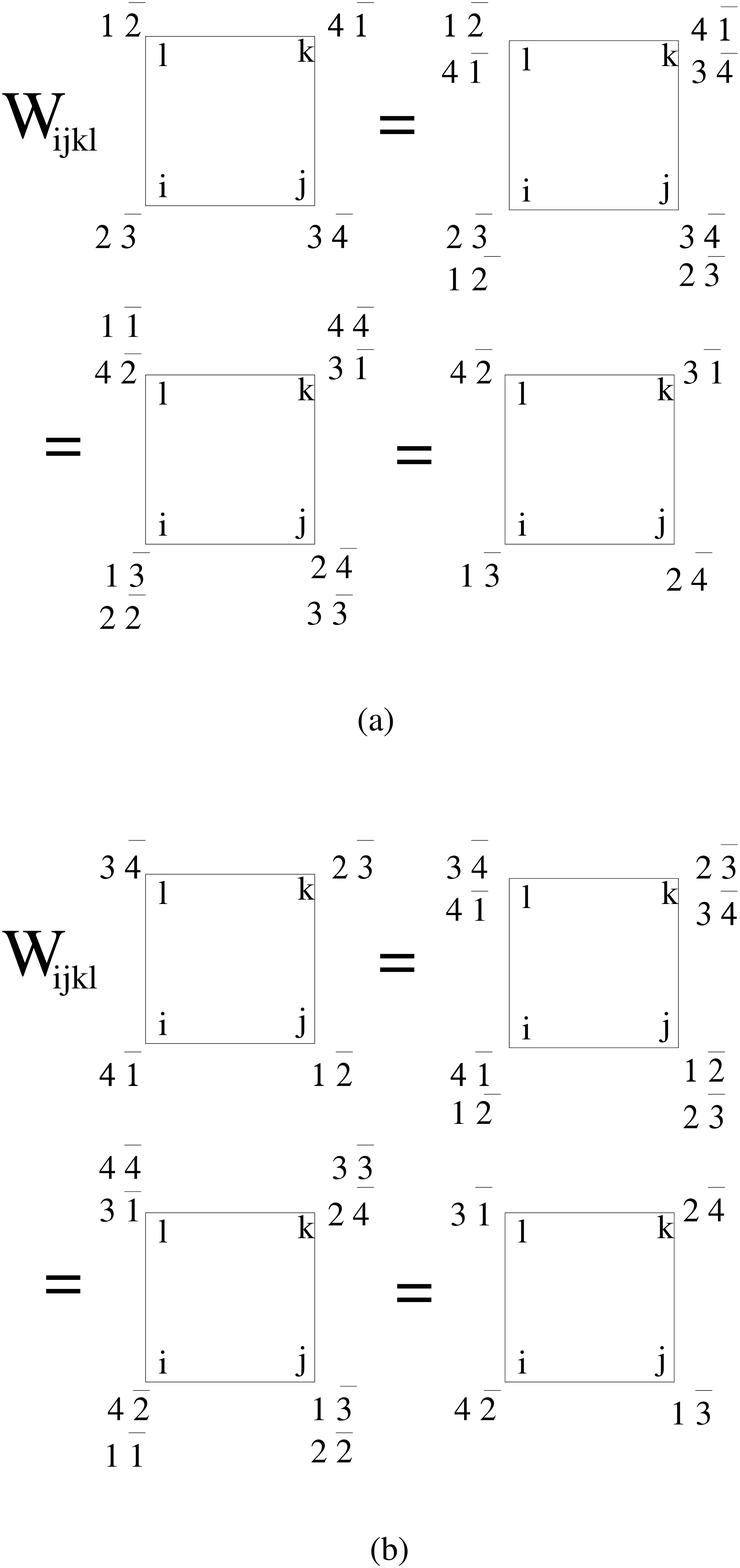}
\caption{
The flavor current caused by the Wilson operator.
$a \bar{b}$ indicates an exciton made of particle in the a-th band
and hole in the b-th band.
The Wilson operator induces the motion of boson in the counter-clockwise direction and anti-boson 
in the clockwise direction.
}
\label{fig:wilson_op}
\end{figure}

What does the Wilson operator do to existing excitons ?
It creates flavor current.
For example if we apply 
$W \sim e^{i(\theta^{12}_i + \theta^{23}_j + \theta^{34}_j + \theta^{41}_l )}$
to the states
\bqa
|\Psi_1> & = & e^{i(\theta^{23}_i + \theta^{34}_j + \theta^{41}_j + \theta^{12}_l )} |0>, \nn
|\Psi_2> & = & e^{i(\theta^{41}_i + \theta^{12}_j + \theta^{23}_j + \theta^{34}_l )} |0>
\eqa
with $|0>$ being a vacuum without any off-diagonal exciton,
we have
\bqa
W  |\Psi_1> & = & e^{i(\theta^{13}_i + \theta^{24}_j + \theta^{31}_j + \theta^{42}_l )} |0>, \nn
W  |\Psi_2> & = & e^{i(\theta^{42}_i + \theta^{13}_j + \theta^{24}_j + \theta^{31}_l )} |0>.
\eqa
Here we used the fact that once $ab$ and $bc$ excitons are on a site 
they decay into $ac$ and $bb$ excitons to lower energy by condensing the diagonal exciton.
Keeping only the off-diagonal excitons we obtain the above result.
This is shown in Fig. \ref{fig:wilson_op}.
The Wilson operator induces flavor current in one direction (Fig. \ref{fig:wilson_op} (a)), 
and anti-flavor current (Fig. \ref{fig:wilson_op} (b)) in the other direction, 
thus inducing net flavor current. 
$W_C^\dagger$ induces the opposite current and
the rotational component of the flavor current can be identified as
\bq
\oint_C j^f_\mu dx^\mu \sim ( W_C - W_C^\dagger ),
\eq
where $j^f$ is the flavor current.

\begin{figure}
        \includegraphics[height=8cm,width=7cm]{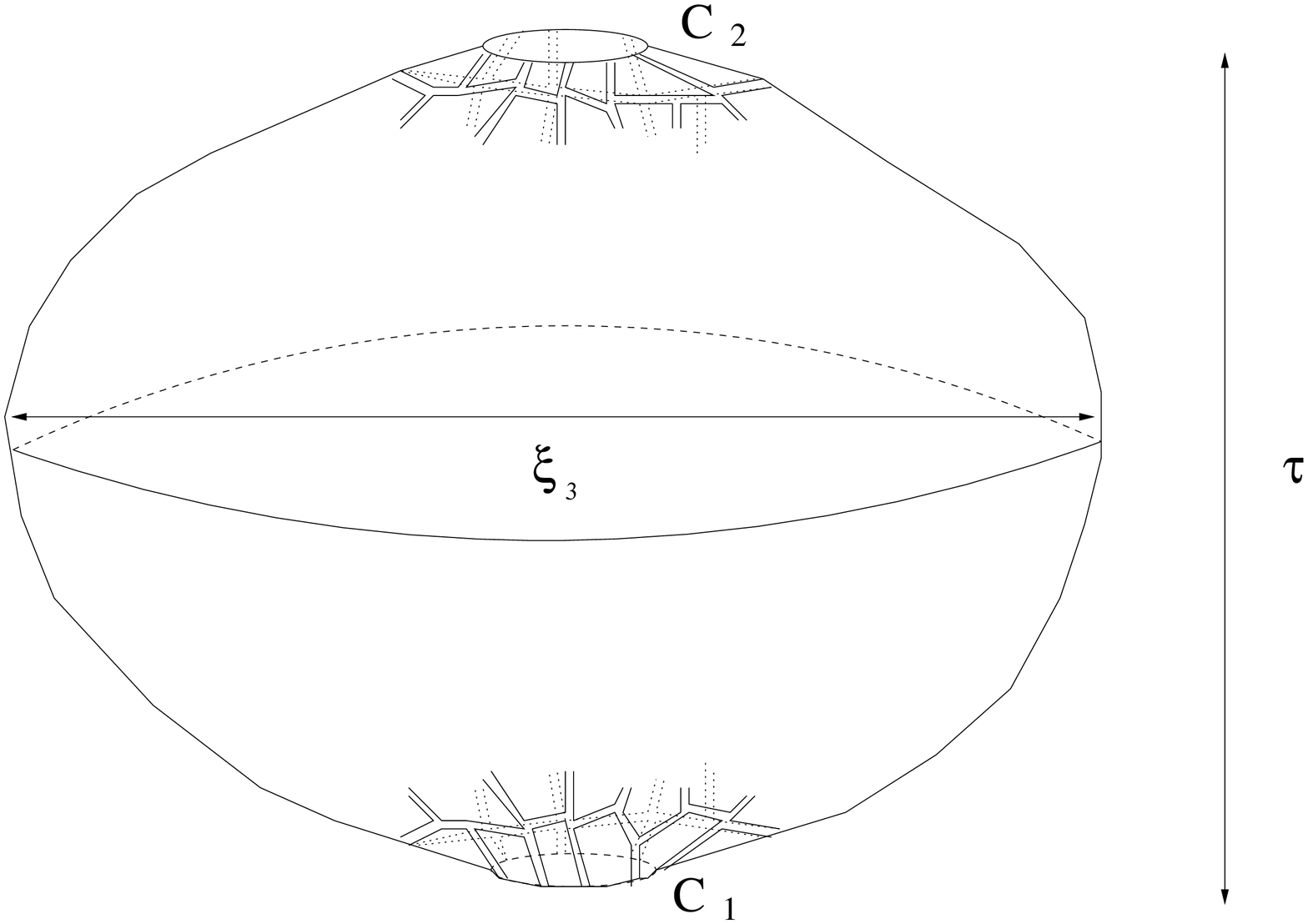}
\caption{
The connected diagram contributing to the correlation function between two Wilson operators.
}
\label{fig:ww}
\end{figure}

With the excitonic expression for the Wilson operator we are readily to
discuss the physical origin of the emergent photon.
In the Coulomb phase there exist infinitely large bubbles of world line web.
This causes long range correlation between the Wilson loop.
Consider the correlator between the fluctuations of two Wilson loops,
\bqa
\left< \delta W_{C_2}^\dagger \delta W_{C_1} \right>,
\eqa
where $\delta W_C = W_C - < W_C >$ and
$C_1$, $C_2$ are two loops separated by $\tau >> \xi_2$ in the imaginary time direction.
Since $W_{C_1}$ ($W_{C_2}^\dagger$) creates (annihilates) world line web, 
the loops $C_1$ and $C_2$ becomes the boundary of the closed surface
as is shown in Fig. \ref{fig:ww}.
Only the connected diagram contribute to the correlator because the background value is subtracted.
In the confining phase with finite $\xi_3$, the contribution of the
connected diagram is exponentially small for $\tau >> \xi_3$.
In the deconfinement phase, there are bubbles of arbitrarily large size
which connect the $C_1$ and $C_2$.
This will leads to power law decay in the correlator, which signifies
the presence of massless mode.
This is the emergent photon in the Coulomb phase.
In this way the emergent photon can be understood in terms of the exciton language. 
In the fractionalized boson picture, the emergent photon is the long-wavelength fluctuation of the flavor currents.
The flavor current in Fig. \ref{fig:wilson_op} can be regarded as if the bosons are really liberated from the anti-bosons.
On the other hand what is really happening is the concerted motion of the neutral excitons 
and the exchange of flavors between excitons on a site.
Therefore the emergent boson should be electromagnetically neutral.

\begin{figure}
        \includegraphics[height=6cm,width=9cm]{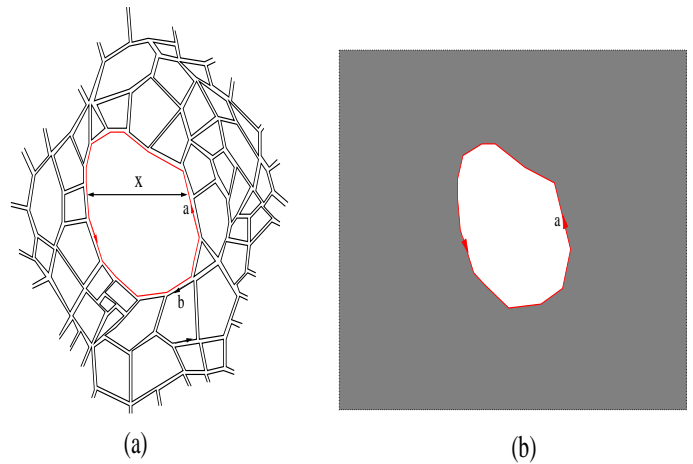}
\caption{
A large loop surrounded by small loops in the exciton world line web (a).
In the length scale comparable to the large loop, the small loops are regarded as smooth 
surface connecting the large loop (b).
}
\label{fig:smooth}
\end{figure}

As $\kappa$ keep increasing $\xi_2$ grows further. 
As seen from Fig. \ref{fig:web}, areas with large $\xi_2$ correspond to punctures
of the bubble.
In the length scale $x \leq \xi_2$ the individual loops bordering the punctures 
on the 2-dimensional surface begin to be noticeable.
The surface is no longer closed surface. 
The individual bosons form the boundary of world sheets and appear as low energy excitations (not necessarily quasiparticle).
For a given length scale $x$, the loops with sizes smaller than $x$ form smooth surface
and are regarded as parts of electric flux line of the U(1) gauge field.
The loops with sizes larger than $x$ are regarded as the world lines of the fractionalized boson.
This is displayed in Fig. \ref{fig:smooth}.
The size of the large loop represents the inverse mass of the fractionalized boson.
A finite loop implies a nonzero mass of the fractionalized boson.
Note that only one flavor of exciton is observable in the large length scale 
because the smooth surface does not carry net flavor.
This is how the fractionalized boson arises.
In a given time slice, the boson and anti-boson
are the end of strings formed by the intersection of the
web surface and the time slice\cite{WEN}.
So the low energy theory becomes the non-compact U(1) gauge theory coupled with gapped bosons.
In the large $\kappa$ limit not only $\xi_3$ but also $\xi_2$ diverges. 
The slave boson itself condenses.
This is the Higgs phase.
Unlike the Coulomb phase the Higgs phase can occur even in the 2+1 dimension.

\subsection{Duality description of fractionalized boson, Ising mode and U(1) gauge field}

\begin{figure}
        \includegraphics[height=5cm,width=5cm]{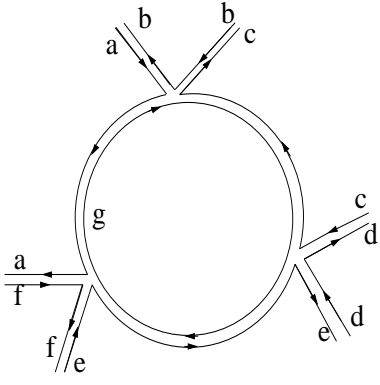}
\caption{
The generation of the sixth order vertex from the quartic order vertices.
}
\label{fig:6th}
\end{figure}

\begin{figure}
        \includegraphics[height=6cm,width=9cm]{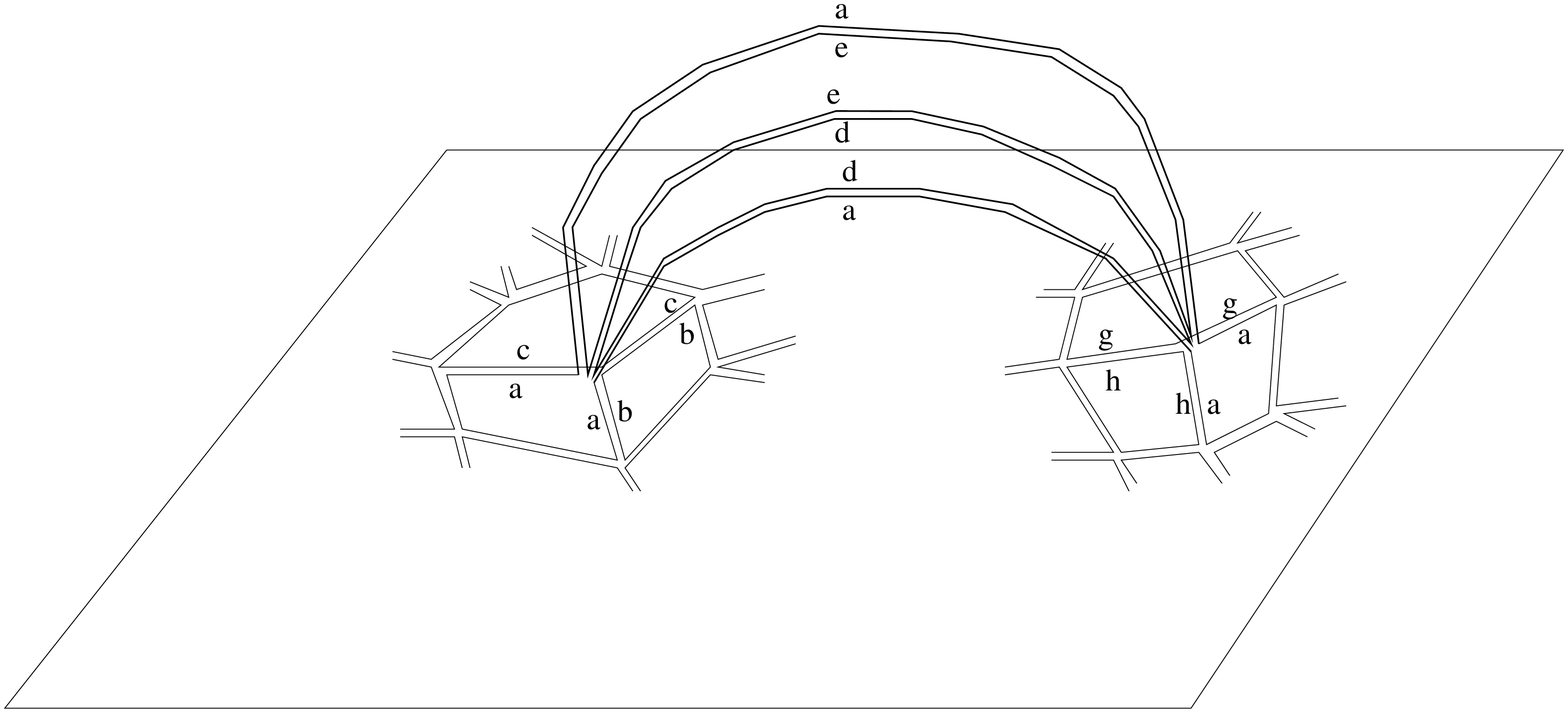}
\caption{
Two sixth order vertices connected by the world lines of three excitons.
}
\label{fig:ising}
\end{figure}

The duality mapping for the Hermitian matrix model 
with $\tilde K_{2n-1} =0$ and  $\tilde K_{2n} <0$ 
is exactly same as the one in the previous section
except for the fact that only vertices of even order are allowed.
Thus we will not repeat the same procedure.
Instead we will discuss the new feature which arises from the absence of vertex of odd order. 
As discussed in the Sec. III. B, there exists extra $Z_2$ degree of freedom
which undergoes the disorder to order phase transition.
How can we interpret the phase transition in the dual representation?
We will show that the phase transition corresponds to a spontaneous generation of the odd order vertex.
To see this, we first note that the presence of quartic vertex implies
the presence of all higher vertices of even order. 
For example, Fig. \ref{fig:6th} shows the generation of a sixth order vertex from the one-loop contribution. 
In the length scale larger than the loop, the sixth order vertex can be regarded as a local interaction.
Two sixth order vertices can effectively act as two third order vertices connected
by world lines of three excitons as is shown in Fig. \ref{fig:ising}.
In the disordered phase of the Ising variable, 
the composite of the three excitons has only short range correlation.
Thus the two third order vertices in Fig. \ref{fig:ising} is linearly confined to each other.
In the ordered phase the world line of the three excitons can be infinitely large.
Thus the third order vertices are unbound.
One can regard the $\tilde K_3$ term as an explicit symmetry breaking term 
for the $Z_2$ symmetry of the exciton Lagrangian.
Then the generation of the third order vertex without bare $\tilde K_3$ is
the spontaneous symmetry breaking of the $Z_2$ symmetry.
In the world line picture it corresponds to the binding unbinding transition of the 
effective third order vertices made of the even order vertices.

\subsection{Duality description of fractionalized fermion, Ising mode and U(1) gauge field}

\begin{figure}
        \includegraphics[height=6cm,width=6cm]{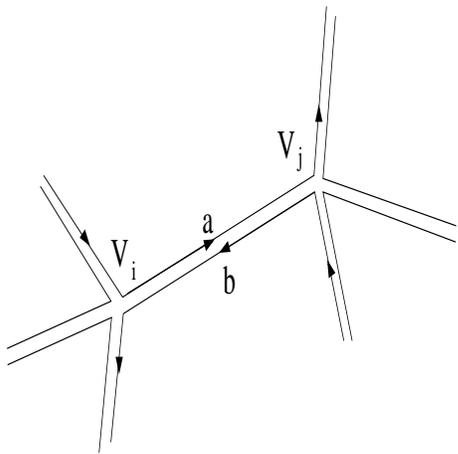}
\caption{
$ab$-exciton is emanating from the vertex $V_i$
and $ba$-exciton, from the vertex $V_j$.
Thus there is one $e^{-i \pi}$ factor associated 
with the exciton world line either from $V_i$ or $V_j$.
}
\label{fig:double_index}
\end{figure}

Here we examine the anti-Hermitian matrix model with $\tilde K_{2n} > 0$.
The duality mapping can be performed in a parallel way to the mapping for the fractionalized boson except for the two differences.
First, the coefficients of the potential energy term has opposite sign. 
Second, the matrix is anti-Hermitian.
These two differences cause the transmutation of the statistics of the slave particle. 
The effective action for the off-diagonal phases is written
\begin{widetext}
\bqa
S & = & 
-\frac{\kappa}{2} \sum_{i} \sum_{\mu} \sum_{a<b} 
\left[ \cos \left( \theta^{ab}(i+\mu) - \theta^{ab}(i) \right)  -1 \right] \nn
&& - \sum_{n \geq 2} \frac{\kappa_{2n}}{2} \sum_{i} \sum_{[a_1,a_2,...,a_{2n}]}
f^{[a_1,a_2,...,a_{2n}]}
\left[
\cos \left( \sum_{l=1}^{2n} \theta^{a_l a_{l+1}}(i)  - \pi   \right) - 1 \right],
\label{dfs1}
\eqa
where $\theta^{ab} = - \theta^{ba} + \pi $  is the phase of the anti-Hermitian matrix.
Other notations are same as those in Sec. V. A.
Note that that there is an extra phase $\pi$ in the potential term accounting for the opposite sign of the coupling constant.

From the Villain approximation and the Hubbard-Stratonovich transformations we obtain the similar partition function to (\ref{db1z}),
\bqa
Z & = &
\sum_{l_{\mu}^{ab}(i), V^{[a_1...a_{2n}]}(i) }   
Q_V
 \exp \Bigl( 
-\frac{1}{\kappa} \sum_{i,\mu,a<b} (l_{\mu}^{ab}(i))^2   
- \sum_{n=2}^{\infty} \sum_{i} \sum_{[a_1,a_2,...,a_{2n}]} \frac{1}{\kappa^{[a_1...a_{2n}]}}  (V^{[a_1 ... a_{2n}]}(i))^2
\Bigr) \times \nn
&& \int D \theta  
\exp \Bigl(
-i \sum_{i,\mu,a<b} l_{\mu}^{ab}(i) \{ \theta^{ab}(i+\mu) - \theta^{ab}(i) \}  
-i \sum_{n=2}^{\infty} \sum_{i} \sum_{[a_1,a_2,...,a_{2n}]} 
  V^{[a_1 ... a_{2n}]}(i) \sum_{l=1}^{2n} \theta^{a_l a_{l+1}}(i)  
\Bigr).  \nn
\eqa
\end{widetext}
The extra phase factor $Q_V$ in the partition function is the consequence of 
the opposite sign in the coupling constant.
It is given by
\bq
Q_V \equiv 
\exp \Bigl(
i \pi \sum_{n}^{\infty} \sum_{i} \sum_{[a_1,a_2,...,a_{2n}]} V^{[a_1 ... a_{2n}]}(i) 
\Bigr)
\eq
and it can be simply written as
\bq
Q_V = e^{i \pi V},
\eq
where $V$ is the total number of vertices in the polygon made of the exciton world lines.
Besides $Q_V$, there is another phase arising from the anti-Hermitivity.
Since $\theta^{ab}$ and $\theta^{ba}$ are not independent fields,
one has to replace  $\theta^{ba}$ into $- \theta^{ab} + \pi$ 
for $a < b$ in integrating over $\theta^{ab}$. 
After integration it leads to the same constraint as (\ref{const})
but with another phase factor, 
\bqa
Q_E & = &
\exp \Bigl(
-i \pi \sum_{n=2}^{\infty} \sum_{i} \sum_{[a_1,a_2,...,a_{2n}]} 
  V^{[a_1 ... a_{2n}]}(i) s^{[a_1 ... a_{2n}]}(i) 
\Bigr), \nn
\eqa
where 
\bq
s^{[a_1 ... a_{2n}]}(i) = \sum_{i=1}^{2n} \frac{1}{2} \left( \frac{a_i - a_{i+1}}{| a_i - a_{i+1} |}  + 1 \right).
\eq
$s^{[a_1 ... a_{2n}]}(i)$ counts the number of emanating exciton world lines attached to the vertex
$V^{[a_1 ... a_{2n}]}(i)$ whose double index is in descending order
(e.g. $s^{[1234]}(i) = 1$ because there is one descending double index $41$ among $12$, $23$, $34$, $41$;
similarly we have $s^{[1243]}(i) = 2$.)
The second extra phase factor, in turn, becomes
\bq
Q_E = e^{-i \pi E},
\eq
where $E$ is the total number of exciton world line segments.
This is because each exciton world line is attached to two vertices and 
the double index is in ascending order in one vertex 
and in descending order in the other vertex
(see Fig. \ref{fig:double_index}).

\begin{figure}
        \includegraphics[height=5cm,width=9cm]{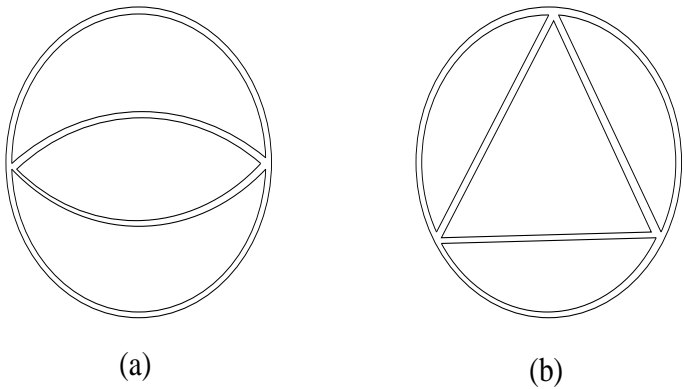}
\caption{
Configurations of exciton world lines
involving (a) $4$ loops and (b) $5$ loops of single line.
}
\label{fig:45loops}
\end{figure}

\begin{widetext}

Finally we have
\bqa
Z & = &
\sum_{l_{\mu}^{ab}(i), V^{[a_1...a_{2n}]}(i) }   
e^{ i\pi(V-E)} 
 \exp \Bigl( 
-\frac{1}{\kappa} \sum_{i,\mu,a<b} (l_{\mu}^{ab}(i))^2    
 - \sum_{n=2}^{\infty} \sum_{i} \sum_{[a_1,a_2,...,a_{2n}]} 
\frac{1}{\kappa^{[a_1...a_{2n}]}}  (V^{[a_1 ... a_{2n}]}(i))^2  
\Bigr) 
\times \nn
&& \left( \Pi_{i,a<b} \delta_{\sum_\mu \Delta_\mu l^{ab}_{\mu}(i) - \sum_{n} \sum_{a_3,..,a_{2n}} (V^{[ab a_3...a_{2n}]} - V^{[ba a_3...a_{2n}]} ), 0}  \right).
\eqa
\end{widetext}
The extra phase factor can be related to the number of faces $F$ of the polygon
because the Euler characteristic is a topological invariant,
\bq
\chi_{Euler} \equiv V - E + F = 2(1-g),
\label{Euler}
\eq
where $g$ is the genus of the surface.
Since $g$ is integer we have
\bqa
e^{i\pi (V-E)} & = & e^{i \pi F}.
\label{vef}
\eqa
On the other hand the number of the faces is same as the number of the loops of single line.
There is extra phase factor $(-1)$ for every loop of single line.
Thus each single line represents the world line of a fermion.
For example the diagram in Fig. \ref{fig:45loops} (a) has $4$ loops 
and Fig. \ref{fig:45loops} (b) has $5$ loops of single line.
They contribute to partition function with the opposite sign.
This is the signature of the emergent fermion in the exciton partition function.
As was mentioned before, the orientedness of the closed surface is important for 
the simple relation (\ref{vef}) 
because the Euler characteristic (\ref{Euler}) can be odd integer 
for non-orientable surface.
It is straightforward to obtain the same result
from the Hermitian matrix model with
$\tilde K_{2n} = (-1)^n |\tilde K_{2n}|$.

The confinement to deconfinement phase transition can be understood as the
divergence of the scale of the world line web as discussed in Sec. V. A.
The difference is that the bubbles made of exciton world lines contribute
to the partition function with oscillating sign as the number of single line loops 
increases.
The oscillating sign also appears in the partition function for the
gauge theory coupled to fermions as seen from 
the phase factor $e^{i\pi n_C}$ in Eq. (\ref{gfz}).
Thus we see that this case corresponds to an 
emergent gauge theory with the opposite sign of the
kinetic energy term (Eq. (\ref{gfs})).

The phase transition for the $Z_2$ variable can also be understood in the
same way as discussed in Sec. V. B.
The difference is that the ordered phase of the Ising variable
is not related to the generation of the usual
$\tilde K_3 tr^{'} \chi^3$ term in (\ref{H}). 
It cancels with its complex conjugate term for the anti-Hermitian matrix.
The spontaneously generated interaction in the ordered phase corresponds to 
a flavor dependent interaction,
\bq
\sum_{a \neq b \neq c} K_3^{abc} \chi^{ab} \chi^{bc} \chi^{ca}
\label{df_K3}
\eq
with $K_3^{abc} = - K_3^{cba}$.
With the flavor dependent coupling constant, 
the complex conjugate term no longer cancel each other
under the anti-Hermitian constraint.
Eq. (\ref{df_K3}) contributes to (\ref{dfs1}) as vertex of third order.
Note that the overall sign of $K_3^{[a,b,c]}$ does not matter
because the third order vertex should occur in pair
because of the constraint between the number of vertices
and the double line segments,
\bq
\sum_n n V_n = 2E.
\eq
Finally, the massless fermion can arise if the size of the loop diverges.

\section{Conclusions}

In this paper we studied an exciton model which supports various fractionalized phases
with either fractionalized boson or fermion along with emergent photon.
Phenomenological matrix model for exciton bose condensate is derived from a microscopic theory.
The statistics and spin of the fractionalized particles are determined dynamically.
The bosonic fractionalized particle is shown to occur more generically.
The fermionic nature of the fractionalized particle is shown to be rooted to 
frustration in the underlying bosonic matrix model.
The fermionic statistics, in turn, is shown to be responsible for the 
generation of spin in the continuum limit in agreement with the spin-statistics theorem.
The fractionalization is explained in two consistent point of views.
One is based on more conventional approach where the fractionalized degrees of freedom
are obtained from decomposition of original exciton field into multiple modes
as required by constraints imposed by strong interaction.
The other approach is based on the world line representation of original exciton.
In the latter approach, a more vivid physical picture for the fractionalization 
is obtained in terms of the original exciton language,
while more quantitative but abstract analysis is given in the first approach.

We close by making some comments on the observational consequences 
of the emergent photon and fractionalized particles in the
Coulomb phase.
The photon is the only gapless mode and will dominate the
specific heat at low temperature, as $T^3$ in $d=3$
and $T^2$ with cross-over to a small gap to $d=2$.
The fractionalization of the exciton will greatly 
affect the spectral properties.
The exciton is not a well defined excitation.
The exciton propagator is a product of two fractionalized particles
in space-time, connected by a strongly fluctuating web, i.e.
interacting via the exchange of photons.
If the coupling to the photon is weak, it can be ignored and the exciton 
spectral function is a convolution of two spectral functions of 
fractionalized particles, and will become extremely broad with
an unusual line-shape.

\section{Acknowledgements}
We thank X.-G. Wen and T. Senthil for illuminating discussions.
This work was supported by the NSF grant DMR-0517222.

\newpage
\renewcommand{\theequation}{A\arabic{equation}}
\setcounter{equation}{0}
\section*{Appendix A. Effective action of exciton condensate and the saddle-point solutions}

The objective of this appendix is two-folded.
The first one is to derive the
effective action (\ref{ls})-(\ref{ls2}) from the
microscopic Hamiltonian (\ref{micro}).
The second one is to present the saddle point solutions for 
the self-consistent equations for the exciton condensate field.

We rearrange the the attractive interaction as
\bqa
&& - \sum_{ab} V^{ab}
\sum_{k,k^{'},q} c^{a*}_{k+q} c^a_k  d^b_{k^{'}+q} d^{b*}_{k^{'}}  \nn
& = &
- \sum_{ab} V^{ab}
\sum_{k,k^{'},q}
d^{b*}_{k} c^a_{k+q} c^{a*}_{k^{'}+q} d^b_{k^{'}},  
\eqa
and introduce Hubbard-Stratonovich fields describing the exciton condensate
to obtain
\bqa
L_{int}
& = &
\sum_{ab}
\sum_q \left(
(V^{ab})^{-1}
|\xi^{ab}_q|^2 
- \frac{\xi^{ab}_q}{\sqrt{v}} \sum_{k^{'}} c^{a*}_{k^{'}+q} d^b_{k^{'}} \right. \nn
&& \left. -\frac{\xi^{ab*}_q}{\sqrt{v}} \sum_{k} d^{b*}_{k} c^a_{k+q} 
\right). 
\eqa
With the anisotropic Coulomb interaction, $V^{ab} = V + V^{'} \delta_{ab}$
and a dimensionless exciton field, $\chi^{ab} = \frac{\xi^{ab}}{V}$,
the effective Lagrangian becomes 
\bqa
L & = &
 V  \sum_q \sum_{ab} |\chi^{ab}_q|^2
- \frac{V V^{'}}{V+V^{'}}  \sum_q \sum_{a} |\chi^{aa}_q|^2 \nn
&& 
+ \sum_k \sum_a 
\Bigl[
     c^{a *}_k  \left(\partial_\tau + \epsilon^c_k - \mu \right) c^a_k  
 + d^{a *}_k  \left(\partial_\tau + \epsilon^d_k - \mu  \right) d^a_k 
\Bigr]  \nn
&& - \frac{V}{\sqrt{v}} \sum_q \sum_{ab} \left(
 \sum_{k} c^{a *}_{k+q} \chi_q^{ab}  d^b_{k} 
+ \sum_{k} d^{b *}_{k} \chi_q^{ab*} c^a_{k+q} 
\right) \nn
&& + U \sum_k \sum_{a} \left(
  c^{a *}_{k}  d^a_{k} 
+ d^{a *}_{k} c^a_{k} 
\right).
\eqa
Absorbing the mixing ($U$) term into $\chi$ by a shift of the exciton field
\bq
\chi^{ab}_q  = \chi^{'ab}_q + \frac{U}{V} \sqrt{v} \delta_{ab} \delta_{q,0},
\eq
we finally obtain the Lagrangian,
\bqa
L & = &
 V  \sum_q \sum_{ab} |\chi^{'ab}_q|^2
- \frac{V V^{'}}{V+V^{'}}  \sum_q \sum_{a} |\chi^{'aa}_q|^2 \nn
&& + \frac{VU}{V+V^{'}} \sqrt{v} \sum_a ( \chi^{'aa}_0 + \chi^{'aa*}_0 ) \nn
&& + \sum_k \sum_a \Bigl[
c^{a *}_k  (\partial_\tau + \epsilon^c_k - \mu ) c^a_k  \nn
&& + d^{a *}_k (\partial_\tau + \epsilon^d_k - \mu  ) d^a_k 
\Bigr]  \nn
&& - \frac{V}{\sqrt{v}} \sum_q \sum_{ab} \left(
 \sum_{k} c^{a *}_{k+q} \chi_q^{'ab}  d^b_{k} 
+ \sum_{k} d^{b *}_{k} \chi_q^{'ab*} c^a_{k+q} 
\right) \nn
&& + \frac{N U^2}{V+V^{'}} v
\eqa
From now on we omit the prime in the $\chi^{'}$ and drop the last constant term.

Now we consider the mean-field solution for the exciton condensate.
We consider a translationally invariant saddle point Ansatz,
\bq
\chi_q^{ab}  =  \sqrt{v} \chi^{ab} \delta_{q,0}
\eq
for the exciton condensate
and solve the self-consistent equations for $\chi^{ab}$,
\bqa
\left(  
 1 - \frac{V^{'}}{V+V^{'}} \delta_{ab}
\right) \chi^{ab}
+ \frac{U}{V+V^{'}} \delta_{ab}   & = & 
\frac{1}{v} \sum_k < d_k^{b\dagger} c_k^a >. \nn
\eqa
Note that the saddle point equation for $\chi^{ab}$ has different form 
from (\ref{order}) because of the rescaling and shift in $\chi^{ab}$.

Here we present the saddle point solutions for $N=3$ obtained in the $20 \times 20$ square lattice.
For simplicity we use the particle-hole symmetric tight-binding dispersion , 
$\epsilon^c_k - \mu = - (\epsilon^d_k - \mu) =  -t ( \cos k_x + \cos k_y - 2 ) + \Delta$ with
$t=1$ and $\Delta = 1$. 
Various solutions are possible depending on the parameters of $V$, $V^{'}$, $U$.
The followings are the saddle point solutions :

\begin{itemize}
\item $V = 10$, $V^{'} = 5$, $U=0$ :

\bqa
\chi^{ab} & = & 
\left(
\begin{array}{ccc}
\chi_d & 0 & 0 \\
0 & \chi_d & 0 \\
0 & 0 & \chi_d \\
\end{array}
\right), 
\eqa
with $\chi_d = 0.68$.

\item $V = 10$, $V^{'} = -5$, $U=0$ :

\bqa
\chi^{ab} & = & 
\left(
\begin{array}{ccc}
0 & \chi_1 & 0 \\
0 & 0 & \chi_1 \\
\chi_1 & 0 & 0 \\
\end{array}
\right), 
\eqa
with $\chi_1 = 0.4$.

\item $V = 10$, $V^{'} = -5$, $U=0.3$ :

\bqa
\chi^{ab} & = & 
\left(
\begin{array}{ccc}
\chi_d & -\chi_1 & \chi_2 \\
-\chi_2 & \chi_d & -\chi_1 \\
\chi_1 & -\chi_2 & \chi_d \\
\end{array}
\right), 
\eqa
with $\chi_d = -0.06$, $\chi_1 = 0.07$ and $\chi_2=0.39$.

\item $V = 10$, $V^{'} = -5$, $U=0.7$ :

\bqa
\chi^{ab} & = & 
\left(
\begin{array}{ccc}
\chi_d & \chi_o & \chi_o \\
\chi_o & \chi_d & \chi_o \\
\chi_o & \chi_o & \chi_d \\
\end{array}
\right), 
\eqa
with $\chi_d = -0.14$ and $\chi_o = 0.27$.

\end{itemize}

The above solutions are locally stable. 
However it is possible that they are not global minima.
One may have to add further interaction term to stabilize the desired saddle point solution.

\renewcommand{\theequation}{B\arabic{equation}}
\setcounter{equation}{0}
\section*{Appendix B. Derivation of Hermitian and anti-Hermitian matrix models}

In this appendix we derive the Hermitian (\ref{H}) and the anti-Hermitian matrix model (\ref{antiH}) from 
the effective action of the exciton condensate (\ref{theaction}).
This is done by replacing the diagonal element of $\chi$ by
the amplitude of the diagonal condensate, $\chi_d$  in (\ref{theaction})
and extract the terms for the off-diagonal elements of $\chi$.
In the quartic interaction
\bqa
K_4 \sum_{a,b,c,d} \chi^*_{ba} \chi_{bc} \chi^*_{dc} \chi_{da},
\eqa
there are $5$ different possibilities of how the diagonal and off-diagonal elements
are mixed depending whether $2$, $1$ or $0$ of the $\chi$ fields are
diagonal.
This leads to the action for the off diagonal phase modes, 
\bqa
L_{K_4} & = & K_4 \chi_d^2 tr^{'} ( \chi \chi  + h.c. ) \nn
&& + 2 K_4 \chi_d tr^{'} ( \chi \chi^\dagger \chi + h.c. ) \nn
&& + K_4 tr^{'} \chi^\dagger \chi \chi^\dagger \chi,
\eqa
where $tr^{'} (A_1 A_2 .. A_n)$ denotes trace for the product of matrices which involves only the off-diagonal element of each matrix, e.g., $tr^{'} AB = \sum_{a \neq b} A_{ab} B_{ba}$.
Similarly, the six-th order interactions also give rise to potential energy 
for the off-diagonal phase modes,
\bqa
L_{K_6} 
& = & 3 K_6 \chi_d^4 tr^{'} ( \chi \chi  + h.c. )
+ K_6 \chi_d^3 tr^{'} ( \chi^3 + h.c. )   \nn
&& + 9 K_6 \chi_d^3 tr^{'} ( \chi \chi^\dagger \chi + h.c. ) 
 + 3 K_6 \chi_d^2 tr^{'} ( \chi^3 \chi^\dagger  + h.c. )  \nn
&&  + 3 K_6 \chi_d^2 tr^{'}  \chi^\dagger \chi^\dagger \chi \chi  
 + 6 K_6 \chi_d^2 tr^{'}  \chi^\dagger \chi \chi^\dagger  \chi   \nn
&& + 3 K_6 \chi_d tr^{'}  ( \chi \chi^\dagger \chi \chi^\dagger  \chi + h.c. )  \nn
&& + K_6 tr^{'}   \chi \chi^\dagger \chi \chi^\dagger  \chi \chi^\dagger, \nn
L_{K_6^{'}}  & = & N K_6^{'} ( \chi_d^2 + (N-1) \chi_o^2 ) tr  \chi^\dagger \chi \chi^\dagger  \chi.
\eqa
$L_{K_6^{'}}$ has the same form as $L_{K_4}$ and this renormalizes $K_4$ into
$K_4^{''} \equiv K_4 +  N K_6^{'} ( \chi_d^2 + (N-1) \chi_o^2 )$.
Summing $L_{K_4}$, $L_{K_6}$ and $L_{K_6^{'}}$ we have
\bqa
L_{K} & = &
( K_4^{''} \chi_d^2 + 3 K_6 \chi_d^4  ) tr^{'} ( \chi \chi  + h.c. ) \nn 
&& +  K_6 \chi_d^3 tr^{'} ( \chi^3 + h.c. )   \nn
&& + ( 2 K_4^{''} \chi_d + 9 K_6 \chi_d^3 ) tr^{'} ( \chi \chi^\dagger \chi + h.c. ) \nn
&& + 3 K_6 \chi_d^2 tr^{'} ( \chi^3 \chi^\dagger  + h.c. ) \nn
&&  + 3 K_6 \chi_d^2 tr^{'}  \chi^\dagger \chi^\dagger \chi \chi  \nn
&&  + ( K_4^{''} + 6 K_6 \chi_d^2 ) tr^{'}  \chi^\dagger \chi \chi^\dagger  \chi   \nn
&& + 3 K_6 \chi_d tr^{'}  ( \chi \chi^\dagger \chi \chi^\dagger  \chi + h.c. )  \nn
&& + K_6 tr^{'}   \chi \chi^\dagger \chi \chi^\dagger  \chi \chi^\dagger.
\label{lk}
\eqa

For $\chi_o << 1$ the second order term of $\chi$ is most important.
We rewrite the second order term in (\ref{lk}),
\bqa
\frac{\tilde K_2}{2} tr^{'} ( \chi \chi  + h.c. ) =  \tilde K_2 tr^{'} \chi_o^2 \sum_{a \neq b} \cos( \theta^{ab} + \theta^{ba} ),
\label{k2}
\eqa
where
\bqa
&& \tilde K_2  \equiv  2 ( K_4^{''} \chi_d^2 + 3 K_6 \chi_d^4 ) \nn
&& =  2 \chi_d^2 \left[ K_4 + N K_6^{'} \left( \chi_d^2 + (N-1) \chi_o^2 \right) + 3 K_6 \chi_d^2 \right].  \nn
\eqa
If $\tilde K_2 < 0$,
the energy of (\ref{k2}) is minimized by
\bq
\theta^{ab}_i = - \theta^{ba}_i.
\eq
This gives mass to the $\frac{N(N-1)}{2}$ symmetric modes with $\theta^{ab} = \theta^{ba}$.
The remaining modes are $\frac{N(N-1)}{2}$ antisymmetric modes with $\theta^{ab} = - \theta^{ba}$.
For $| \tilde K_2 | \chi_o^2 >> 1$, the symmetric modes are frozen and 
$\chi$ becomes Hermitian matrix  
($\chi^\dagger = \chi$).
If $\tilde K_2 > 0$,
the energy of (\ref{k2}) is minimized by
\bq
\theta^{ab}_i = - \theta^{ba}_i + \pi
\eq
and $\chi$ becomes anti-Hermitian matrix for $\tilde K_2 \chi_o^2 >> 1$ 
($\chi^\dagger = - \chi$).
These Hermitivity or the anti-Hermitivity condition converts
the dynamics of the off diagonal exciton into relativistic one.
To see this we rewrite the first term in Eq. (\ref{ds}) as
\bqa
S_\tau & = &
- \frac{K}{2}
\sum_{i}  
tr^{'} \chi_{i+ \Delta \tau}^{\dagger} \chi_{i} \nn
& = & 
- \frac{K}{4}
\sum_{i}  
tr^{'} (
\chi_{i+ \Delta \tau}^{\dagger} \chi_{i} 
+ \chi_{i+ \Delta \tau} \chi_{i}^\dagger 
),
\eqa
where we used the Hermitivity or anti-Hermitivity condition
in the last line.
From this, we obtain the Hermitian matrix model,
\bqa
S  & = &
- \frac{K}{4} \sum_{<i,j>} tr^{'} ( \chi_i^\dagger \chi_j + h.c.) 
 + \sum_{n \geq 3} \tilde K_n \sum_i tr^{'} \chi_i^n, \nn
\label{sb}
\eqa
with
\bqa
\tilde K_3 & = & 2 [ 2 K_4^{''} \chi_d + 10 K_6 \chi_d^3 ] ,\nn
\tilde K_4 & = & K_4^{''} + 15 K_6 \chi_d^2, \nn
\tilde K_5 & = & 6 K_6 \chi_d, \nn 
\tilde K_6 & = & K_6
\eqa
for $\tilde K_2 < 0$, and
the anti-Hermitian matrix model,
\bqa
S  & = &
- \frac{K}{4} \sum_{<i,j>} tr^{'} ( \chi_i^\dagger \chi_j + h.c.) 
 + \sum_{n \geq 2} \tilde K_{2n} \sum_i tr^{'} \chi_i^{2n}, \nn
\label{sf}
\eqa
with
\bqa
\tilde K_4 & = & K_4^{''} + 3 K_6 \chi_d^2, \nn
\tilde K_6 & = &  - K_6
\eqa
for $\tilde K_2 > 0$.
Here $<i,j>$ is the nearest neighbor in the D-dimensional (hyper) cubic lattice.
$\chi$ is Hermitian (anti-Hermitian) matrix in Eq. (\ref{sb}) ((\ref{sf})). 
Note that all odd order terms vanish in the anti-Hermitian matrix model
because of the anti-Hermitivity, e.g., 
\bq
tr^{'} ( \chi^\dagger \chi \chi^\dagger + \chi \chi^\dagger \chi   ) 
= tr^{'} ( \chi^\dagger \chi \chi^\dagger - \chi^\dagger \chi \chi^\dagger   )  = 0.
\eq


\begin{thebibliography}{27}
\bibitem{COUNTING}
One can prove this by the method of induction.
It is true for $N=3$.
Suppose it is also true for $N \leq M$.
For $N = M+1$, we have $\frac{(M-1)(M-2)}{2}$
constraints which do not involve $\theta^{M+1,a}$.
There are $M-1$ additional constraints which involves $\theta^{M+1,a}$.
They are $\theta^{M+1,1} + \theta^{1,a} + \theta^{a,M+1} = 0$
with $a=2,...,M$.
( $\theta^{M+1,a} + \theta^{a,b} + \theta^{b,M+1} = 0$
with $a,b >1$ is linear combination of above constraints
and does not count. )
Finally we have $\frac{M(M-1)}{2}$ constraints for $N= M+1$.
\bibitem{GRASSMANN}
The formula for change of variable in integration over Grassmann variable
is $\int d\xi f(\xi) = \int \frac{ d \xi}{a} f(a \xi)$.
\bibitem{8COMP}
If we had allowed the $\pi$-momentum mode in the imaginary time direction
we would have obtained $8$-component Dirac fermion\cite{SUSSKIND}.
In our case, we excluded the high momentum mode in the $z$-direction
because the massless $\pi$-mode is an artifact of the
discretization of the imaginary time,
$|\partial_\tau X |^2 \approx \left( X_{i+\tau}^* X_i + c.c. \right) $.
In the original continuous time action the $\pi$-mode is gapped.
\end{thebibliography}
\end{document}